\documentclass[fdp,a4paper,fleqn%
]{w-art}
\usepackage{times,cite,w-thm}
\theoremstyle{plain}

\theoremstyle{definition}

\usepackage[]{graphicx}
\usepackage{amssymb,amsmath}
\newcommand{\dst}{\displaystyle}
\newcommand{\lsm}{L$\sigma$M}

\newcommand{\qllsm}{QLL$\sigma$M}
\newcommand{\qllsms}{\mbox{\scriptsize QLL$\sigma$M}}
\newcommand{\sutwo}{$SU(2)$}

\newcommand{\lagr}{{\mathcal L}}
\newcommand{\cons}{\mbox{\scriptsize con}}
\newcommand{\cls}{\mbox{\scriptsize CL}}

\newcommand{\ems}{\mbox{\scriptsize e.m.}}
\newcommand{\exps}{\mbox{\scriptsize exp}}
\newcommand{\mhat}{\hat{m}}
\newcommand{\mhatcon}{\hat{m}_{\mbox{\scriptsize con}}}

\newcommand{\mhcur}{\hat{m}_{\mbox{\scriptsize cur}}}
\newcommand{\mscur}{m_{s,\mbox{\scriptsize cur}}}
\newcommand{\mscon}{m_{s,\mbox{\scriptsize con}}}

\newcommand{\scur}{\mbox{\scriptsize cur}}
\newcommand{\scon}{\mbox{\scriptsize con}}
\newcommand{\mdyn}{\hat{m}_{\mbox{\scriptsize dyn}}}

\newcommand{\back}[1]{\hspace*{-#1 pt}}
\newcommand{\pslash}{\not\back{2.15}p}

\newcommand{\dbar}{\mathchar'26\mkern-12mu d}
\newcommand{\dfp}{d^4\:\!\!p}
\newcommand{\dbarfp}{\dbar\;\!^4\:\!\!p}
\newcommand{\dbarfpt}{\dbar^{\,4}\:\!\!p}

\newcommand{\be}{\begin{equation}}
\newcommand{\ee}{\end{equation}}
\newcommand{\bea}{\begin{eqnarray}}
\newcommand{\eea}{\end{eqnarray}}

\newcommand{\bsm}[1]{\boldsymbol{#1}}
\begin{document}
\DOIsuffix{theDOIsuffix}
\Volume{XX}
\Month{XX}
\Year{201X}
\pagespan{1}{}
\Receiveddate{XXXX}
\Reviseddate{XXXX}
\Accepteddate{XXXX}
\Dateposted{XXXX}



\title[The Quark-Level Linear $\sigma$ Model]
      {The Quark-Level Linear $\sigma$ Model}


\author[M. D. Scadron]{Michael D. Scadron\inst{1}
  \footnote{E-mail:~\textsf{scadron@physics.arizona.edu}}}
\address[\inst{1}]{Physics Department, University of Arizona, Tucson,
AZ 85721, USA}
\author[G. Rupp]{George Rupp\inst{2}%
  \footnote{Corresponding author\quad E-mail:~\textsf{george@ist.utl.pt},
            Phone: +351\,218\,419\,103,
            Fax: +351\,218\,419\,143}}
\address[\inst{2}]{Centro de F\'{\i}sica das Interac\c{c}\~{o}es Fundamentais,
Instituto Superior T\'{e}cnico, Universidade de Lisboa,
P-1049-001 Lisboa, Portugal}
\author[R. Delbourgo]{Robert Delbourgo\inst{3}
  \footnote{E-mail:~\textsf{Bob.Delbourgo@utas.edu.au}}}
\address[\inst{3}]{School of Mathematics and Physics, University
of Tasmania \\ GPO Box 252-21, Hobart 7001, Australia}
\begin{abstract}
This review of the quark-level linear $\sigma$ model (\qllsm) is based upon
the dynamical realization of the pseudoscalar and scalar mesons as
a linear representation of $SU(2)\times SU(2)$ chiral symmetry, with the
symmetry weakly broken by current quark masses. In its simplest $SU(2)$
incarnation, with two non-strange quark flavors and three colors, this 
nonperturbative theory, which can be selfconsistently bootstrapped in loop
order, is shown to accurately reproduce a  host of low-energy observables
with only one parameter, namely the pion decay constant $f_\pi$.
Extending the scheme to $SU(3)$ by including
the strange quark, equally good results are obtained for many strong,
electromagnetic, and weak processes just with two extra constants, viz.\
$f_K$ and $\langle\pi |H_{\mbox{\scriptsize weak}}|K\rangle$.
Links are made with the vector-meson-dominance model, the BCS theory
of superconductivity, and chiral-symmetry restoration at high temperature.
Finally, these ideas are cautiously generalized to the electroweak sector,
including the $W, Z$, and Higgs bosons, and also to CP violation.
\end{abstract}
%
\maketitle                   
\section{Introduction}
\label{secintroduction}

The magnitude of the strong interaction between the had\-rons precludes the use
of perturbation theory (PT) and this has been understood for a very long time.
Only in the asymptotic high-energy regime, where the QCD coupling becomes
logarithmically small, does it make any sense to use PT, and then only for the
interactions involving gluons with their effective coupling. Thus it has been
the goal of particle physicists to use nonperturbative schemes in order to
tackle, with any semblance of reliability or conviction, the low-energy
features of hadronic interactions. The sense of this approach is highlighted
by the fact that the current quark masses are so much smaller than the
constituent quark masses within the hadron, so that the extra mass is provided
by the cloud of mesons and gluons which comprise the sum total.
Foremost amongst these nonperturbative approaches has been the application of
spontaneously broken chiral symmetry, accompanied by current algebra. For a
good while the nonlinear realization of chiral symmetry at zero energy was
used, together with an expansion in powers of momentum in order to get away
from that particular limit. Unfortunately this has led to a plethora of
expansion parameters and it blunts the use of the nonlinear theory
predictions. 

However it has also been found that the linear realization of chiral symmetry
at the quark level is an alternative way of handling the low-energy properties
of hadrons, without abandoning spontaneous breaking concepts introduced by
Nambu \cite{PRL4p380:qllsm}. In particular the Gell-Mann--L\'{e}vy model
\cite{NC16p705:qllsm,AFFR73:qllsm}, constrained by a vanishing of the
renormalization constants of the mesons (which makes them composite states) is
extremely predictive for a host of observed phenomena. Indeed it turns out that
for pionic interactions essentially {\em every} \/low energy feature is
determined completely just by one scale, the pion weak decay constant
$f_{\pi}\simeq 92.2$ MeV. In the chiral limit, where the pion mass vanishes,
all the other constants are totally fixed. Thus with just three colors, one
can determine that
\begin{itemize}
 \item the pion-quark coupling is $g=2\pi/\sqrt{3}$;
 \item the quartic pion-pion interaction is $\lambda = 2g^2=8\pi^2/3$;
 \item the constituent nonstrange quark mass is $\hat{m} = f_\pi g \simeq 335$ MeV;
 \item the sigma meson partner to the pion has a mass
$m_{\sigma} = 2\mhat \simeq 670$ MeV.
\end{itemize}
It is even possible to extrapolate away from the chiral limit, by allowing for
small current quark masses (which mar the chiral symmetry slightly) and
thereby determine the deviations. 

All this is explained in detail in Sects.~1, 2, and 3. There, as in succeeding
sections, we compare the predictions of the quark-level linear $\sigma$ model
(\qllsm) with other methods, based on other premises. In Sec.~4 we revisit the
compositeness condition ($Z = 0$), and show how this can be used to set a
demarcation scale between scalar and vector mesons. The importance of chiral
cancellations is reviewed in Sec.~5, as it explains the vanishing of certain
amplitudes, which might otherwise be quite large. This also affects the
$\pi$-$N$ so-called sigma term associated with scattering lengths, treated in
Sec.~6. Section~7 is devoted to the pion charge radius, which is again fully
determined in terms of $f_{\pi}$ as
$r_{\pi} = \hbar c\sqrt{3}/2\pi f_{\pi} \simeq 0.61$ fm, and can be contrasted
with values obtained through the vector dominance model, incidentally
explaining the value of the $\rho$-$\pi$-$\pi$ coupling constant as well. The
breakdown of chiral symmetry at higher temperatures is considered next
(Sec.~8), and this occurs at a critical temperature of $2f_{\pi}$; comparisons
with the BCS (Bardeen, Cooper, Schrieffer \cite{PR108p1175:qllsm}) and NJL
(Nambu--Jona-Lasinio) \cite{PR122p345:qllsm}) models are described there, too.

In Sec.~9, we review the extension to $SU(3)$ by inclusion of the strange
quark mass. Now the ratio $f_K/f_{\pi}$ fixes the constituent strange quark
mass to be about 470 MeV, but its couplings to the quarks remain the same as
the pion's. Furthermore, the $\kappa$ analogue of the $\sigma$ meson is
estimated to be about 797 MeV in mass. This is consonant with equal-mass
splitting laws between scalar and pseudoscalar mesons, and in the process
we review the mixing-angle parameters. Section~10 covers e.m.\ decay
rates, as they constitute clean tests of all that has gone before and 
generally fit the data very well, including the isoscalar scalar
$f_0(500)$ \cite{PLB667p1:qllsm} meson. The other light scalar isoscalar
$f_0(980)$, as well as its isovector partner $a_0(980)$, are dealt with in more
detail in Sec.~11, in particular concerning their strong decays.
Sections~12 and 13 are devoted to weak decays, which are governed by one new
scale: the  $K,\bar{K}$ matrix element of the weak Hamiltonian $H_w$, or
equivalently the transition element $\langle \pi|H_w|K \rangle$. Ramifications
of these ideas to other weak decays are also treated. The e.m.\ form
factors of mesons, their role in certain weak decays, as well as the estimation
of meson polarizabilities form the subject of Sec.~14. In Sec.~15 we establish
a link between the critical temperature, mentioned in Sec.~8, and the BCS
theory of superconductivity with its characteristic energy gap.

Section 16 makes an analogy between the \qllsm\ and the standard electroweak
model. In that picture the Higgs boson is regarded largely as a top-antitop
scalar bound state with a mass of about 315~GeV, set by a weak decay constant
of $f_w \simeq 246$ GeV. This picture is consonant with a weak KSRF
(Kawarabayashi, Suzuki, Riazuddin, Fayyazuddin)
\cite{PRL16p255:qllsm,PRL16p255_2:qllsm} relation and the observed masses of
the weak vector bosons plus the weak mixing angle. Possible implications of 
recently observed \cite{PLB716p1:qllsm,PLB716p30:qllsm} Higgs-like signals at 
the large hadron collider (LHC) of CERN are discussed as well. We conclude this
review in Sec.~17 by an analysis of CP violation, as supposed to arise from a
nonstandard $WW\gamma$ vertex.

\section{$\bsm{SU(2)}$ QLL$\bsm{\sigma}$M}
\label{secqllsm}

First we state the \sutwo\ quark-level linear $\sigma$ model (\qllsm) Lagrangian
density, with interacting part
\be
\lagr^{\mbox{\scriptsize\lsm}}_{\mbox{\scriptsize int}}\;=\;g\bar{\psi}(\sigma+
i\gamma_5\vec{\tau}\cdot\vec{\pi})\psi + g'\sigma(\sigma^2+\vec{\pi}^2) - 
\frac{\lambda}{4}(\sigma^2+\vec{\pi}^2)^2 \; ,
\label{lagrangian}
\ee
with the chiral-limiting (CL) pion-quark and meson-meson couplings
\be
g\;=\;\frac{m_q}{f_\pi} \;\;\; , \;\;\;
g'\;=\;\frac{m^2_\sigma}{2f_\pi}=\lambda f_\pi \; .
\label{couplings}
\ee
This \qllsm\ is in the spirit of the original Gell-Mann--L\'{e}vy \lsm\
\cite{NC16p705:qllsm,AFFR73:qllsm}, but for quarks
\cite{PL8p214:qllsm,PL8p214_2:qllsm} rather than for nucleon fermions,
and also with Nambu--Goldstone
\cite{PRL4p380:qllsm,NC19p154:qllsm,NC19p154_2:qllsm} pseudoscalar pions,
having vanishing mass in the chiral limit, i.e., $m_\pi^{\cls}=0$.
Note, however, that the nonstrange pion and sigma mesons in
Eq.~(\ref{lagrangian}) are quantum fields which both
vanish in the CL. Such a vanishing does not occur in
\cite{NC16p705:qllsm,AFFR73:qllsm}
for spin-$1/2$ nucleons, in contrast with the present \qllsm\ scheme.
Furthermore, in the chiral limit the nonstrange constituent quark mass
$\mhatcon=(m_u+m_d)/2$ is half the mass of the $\sigma$ meson, i.e.,
\be
m_\sigma^{\cls} \; = \; 2 \mhatcon^{\cls} \; ,
\label{mhat}
\ee
which is valid for both the \qllsm\ of Eq.~(\ref{lagrangian}) and the nonlinear
NJL \cite{PR122p345:qllsm} model. The Goldberger--Treiman
relation (GTR) \cite{PR110p1478:qllsm} for the \qllsm\ reads
 $\mhatcon=f_\pi g$, which should be compared with the GTR for nucleons, viz.\
$g_Am_N=f_\pi g_{\pi NN}$, but now at the quark level, with $g_A$=1 for
constituent quarks \cite{PRL65p1181:qllsm}.

\begin{figure}[ht]
\begin{tabular}{ccc}
\includegraphics[scale=0.6]{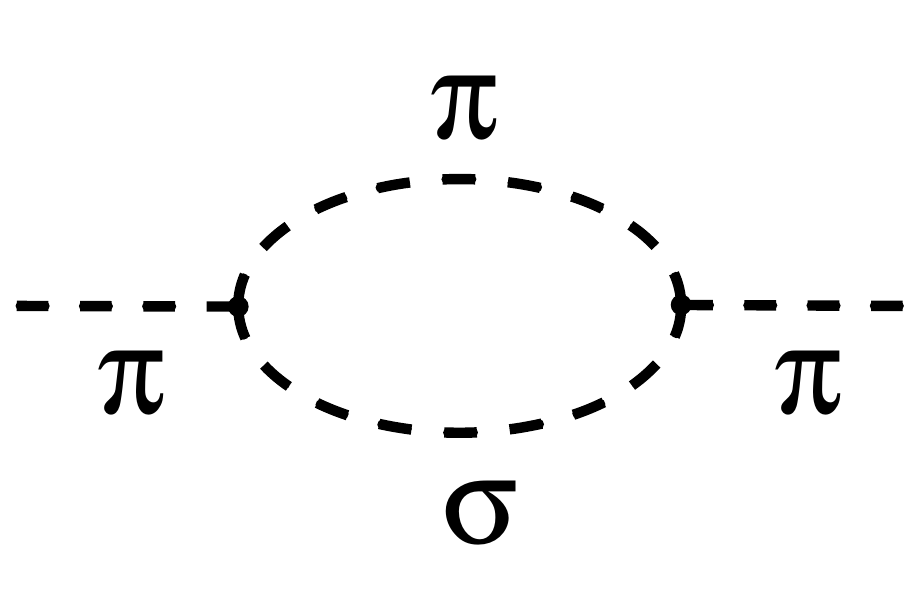} 
& 
\hspace*{8mm} 
\includegraphics[scale=0.6]{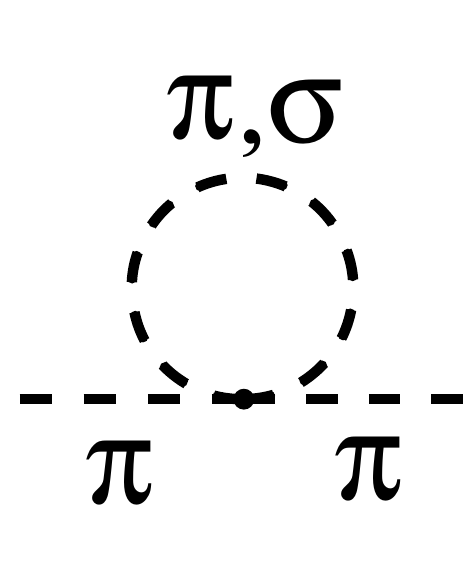}
& \hspace*{8mm} 
\includegraphics[scale=0.6]{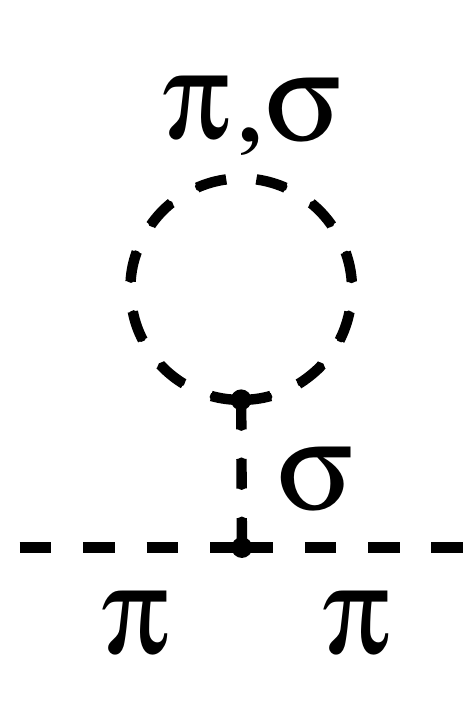} 
\end{tabular}
\mbox{ } \\[-4mm]
\caption{Pion-selfenergy meson-loop graphs. Left: bubble; middle:
``snail''; right: tadpole.}
\label{pionmeson}
\end{figure}

Next we follow \cite{MPLA10p251:qllsm,JPG24p1:qllsm}, and compute the
 pseudoscalar pion mass, which vanishes in the CL, via the selfenergy graphs of
Figs.~\ref{pionquark} and \ref{pionmeson}.
\begin{vchfigure}[h]
\begin{tabular}{cc}
\includegraphics[scale=0.6]{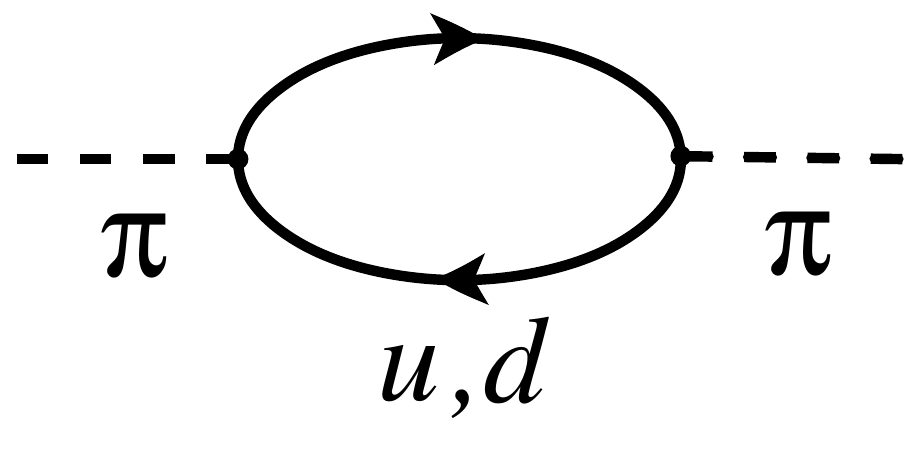} &
\hspace*{15mm}
\includegraphics[scale=0.6]{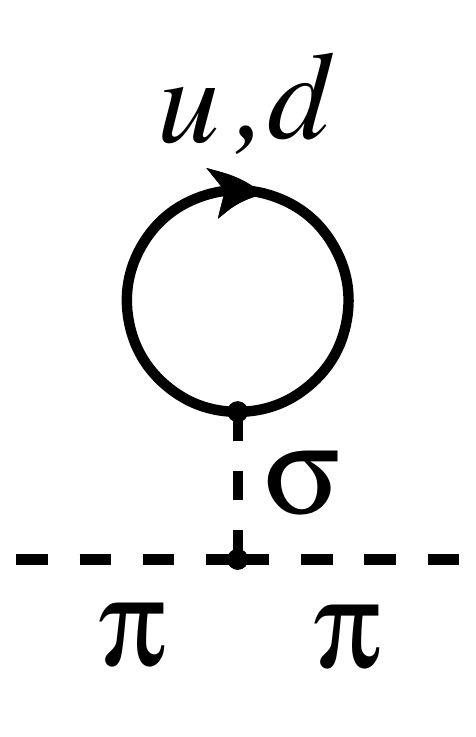} 
\end{tabular}
\mbox{ } \\[-3mm]
\vchcaption{Pion-selfenergy quark-loop graphs. Left: bubble; right: $\sigma$
tadpole.}
\label{pionquark}
\end{vchfigure}
The quark loops (QL) of Fig.~\ref{pionquark} give a pion mass squared
\be
m^2_{\pi,\mbox{\scriptsize QL}} \; = \;
-i8N_cg^2\left(1-\frac{2g'f_\pi}{m^2_\sigma}\right)
\int\frac{\dbarfp}{p^2-m^2_q} \; = \; 0 \; ,
\label{qls}
\ee
with $\dbarfpt\equiv\dfp/(2\pi)^4$. Now,
$m^2_{\pi,\mbox{\scriptsize QL}}$ vanishes identically in the CL, since then
$g'=m^2_\sigma/2f_\pi$. As for the meson-loop (ML) graphs in
Fig.~\ref{pionmeson},
we first invoke the partial-fraction identity
\be
\frac{g'^2}{(p^2-m^2_\sigma)(p^2-m^2_\pi)} \; = \; \frac{\lambda}{2}
\left[\frac{1}{p^2-m^2_\sigma} - \frac{1}{p^2-m^2_\pi}\right] \; ,
\label{partial}
\ee
using $g'^2=\lambda(m^2_\sigma-m^2_\pi)/2$. Then, the contributions from the
diagrams in Fig.~\ref{pionmeson} become
\be
m^2_{\pi,\mbox{\scriptsize ML}} \; = \; i(-2\lambda+5\lambda-3\lambda)
\int\frac{\dbarfp}{p^2-m^2_\pi} \; + \; i(2\lambda+\lambda-
3\lambda)\int\frac{\dbarfp}{p^2-m^2_\sigma} \; = \; 0 \; .
\label{mls}
\ee
Note that the coefficients of the quadratically divergent graphs in
Fig.~\ref{pionmeson} vanish \em identically. \em Combining Eqs.~(\ref{qls})
and (\ref{mls}) generates the vanishing Nambu-Goldstone pion mass in the CL
\be
m^2_\pi \; = \; 0\,|_{\mbox{\scriptsize quark loops}} \; + \;
                0\,|_{\mbox{\scriptsize $\pi$ loops}} \; + \;
                0\,|_{\mbox{\scriptsize $\sigma$ loops}} \; = \; 0 \; .
\label{zero}
\ee
Such is the subtle beauty of chiral symmetry!

Now we determine the $\pi^0$ decay constant $f_\pi$. The latest data in the
Particle Data Group (PDG) \cite{PLB667p1:qllsm} tables give\footnote
{Note that the PDG quotes $f_{\pi^\pm}$, which by definition is larger by a
factor $\sqrt{2}$.}
\be
f_\pi \; = \; (92.21\pm0.15)\;\mbox{MeV} \; .
\label{fpi}
\ee
Also, the pion-quark coupling is \cite{PRL53p1129:qllsm,PRL53p1129_2:qllsm}
\be
g_{\pi qq} \; = \; \frac{2\pi}{\sqrt{N_c}} \; \simeq \; 3.6276 \;,
\label{gpiqq}
\ee
a value to be reconfirmed shortly, for $N_c=3$ colors. Thus, the nonstrange
constituent quark mass found via the GTR becomes
\be
\mhatcon \; = \; f_\pi g_{\pi qq} \; = \; (92.21\:\mbox{MeV})\,
\frac{2\pi}{\sqrt{3}} \; \simeq \; 334.5\;\mbox{MeV} \;.
\label{mhatfg}
\ee
This value is slightly less than a quick estimate resulting from combining the
proton mass with its magnetic moment, i.e., for $\mhatcon=(m_u+m_d)_{\scon}/2$,
\be
\mhatcon \; \simeq \; \frac{m_p}{\mu_p} \; \simeq \;
\frac{938.27\:\mbox{MeV}}{2.7928} \; \simeq \; 336.0\;\mbox{MeV} \;.
\label{mhatmag}
\ee
A more accurate computation from the proton magnetic moment, which takes into
account a 4 MeV mass difference between the down and the up quark, yields the
prediction \cite{JPG32p735:qllsm}
\be
\mhatcon \; \simeq \; 337.5 \;\mbox{MeV} \;,
\label{nrqm}
\ee
which is just a little bit higher than in Eqs.~(\ref{mhatfg},\ref{mhatmag}).

Finally, we work in the CL to generate $f_\pi^{\cls}$ via a once-subtracted
dispersion relation \cite{RPP44p213:qllsm,RPP44p213_2:qllsm} (involving no
arbitrary parameters as in chiral perturbation theory):
\be
f_\pi-f_\pi^{\cls} \; = \; \frac{m^2_\pi}{\pi}\int_0^\infty
\frac{\Im\mbox{m}\,f_\pi(q^2)\,dq^2}{q^2(q^2-m^2_\pi)} \; = \;
\frac{m^2_\pi}{8\pi^2f_\pi} \; .
\label{fpiclfpi}
\ee
With $f_\pi=92.21$ MeV and an average pion mass $m_\pi\simeq137$~MeV, we thus
find
\be
1-\frac{f_\pi^{\cls}}{f_\pi} \; = \; \frac{m^2_\pi}{8\pi^2f_\pi^2} \; \simeq \;
2.8\% \; ,
\label{dfpi}
\ee
which in turn predicts
\be
f_\pi^{\cls} \; = \; f_\pi(1-0.028) \; \simeq \; 89.63\;\mbox{MeV} \; .
\label{fpicl}
\ee
Then, via the GTR, we get
\be
\mhatcon^{\cls}\;  = \; f_\pi^{\cls} g_{\pi qq} \; \simeq  \;
(89.63\;\mbox{MeV})\,\frac{2\pi}{\sqrt{3}} \; \simeq \; 325.1 \;\mbox{MeV}\;.
\label{mconcl}
\ee

\section{Dynamically generating the $\bsm{SU(2)}$ QLL$\bsm{\sigma}$M}
\label{secdynamical}

Following \cite{MPLA10p251:qllsm,JPG24p1:qllsm}, we first compute the
 nonstrange quark loop in Fig.~\ref{quarkloop},
\begin{vchfigure}[h]
\hspace*{1.0cm}
\includegraphics[scale=0.6]{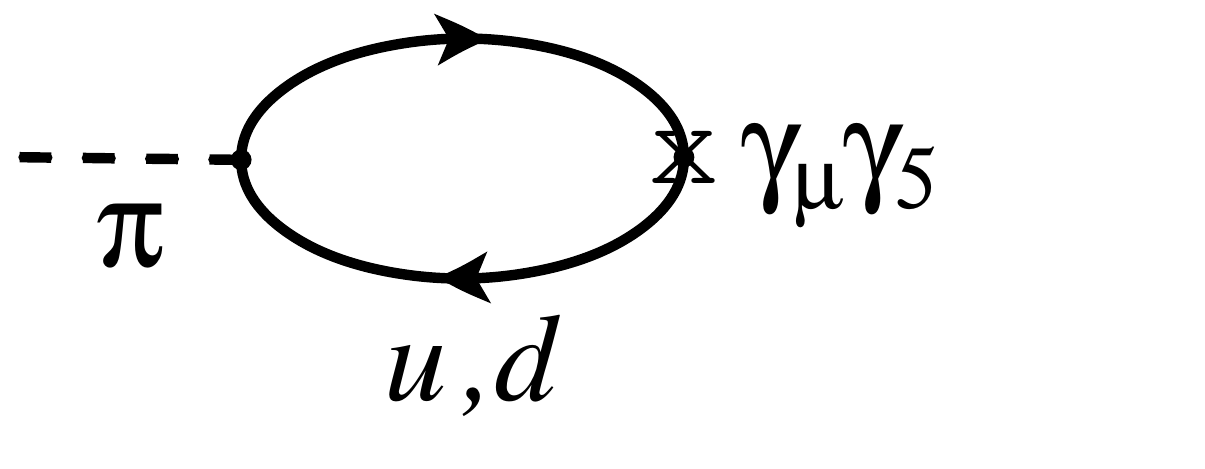} 
\mbox{ } \\[-6mm]
\vchcaption{Pion nonstrange quark loop for the LDGE in Eq.~(\ref{ldge}).}
\label{quarkloop}
\end{vchfigure}
leading to the log-divergent gap equation (LDGE) 
\be
1 \; = \; -i4N_cg^2\int\frac{\dbarfp}{(p^2-m^2_q)^2} \; ,
\label{ldge}
\ee
due to the quark-loop integral for the neutral pion decay constant
combined with the quark-level GTR (\ref{mhatfg}).
This LDGE also holds for the nonlinear NJL scheme in
\cite{PR122p345:qllsm},
and leads to many low-energy theorems \cite{PRD42p941:qllsm}. Furthermore,
 the quark tadpole graph of Fig.~\ref{quarktadpole}
\begin{vchfigure}[h]
\includegraphics[scale=0.6]{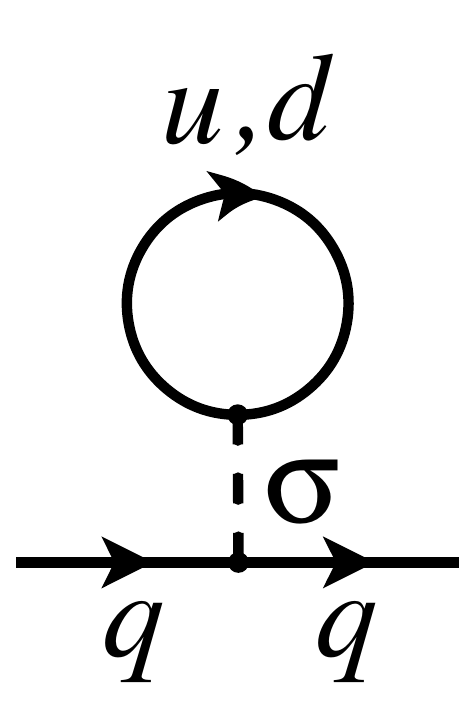} 
\mbox{ } \\[-5mm]
\vchcaption{Quark-selfenergy tadpole graph.}
\label{quarktadpole}
\end{vchfigure}
generates a counter-term mass gap
\be
m_q \; = \; -\frac{8iN_cg^2}{m^2_\sigma}\int\frac{\dbarfp}{p^2-m_q^2}m_q \; .
\label{tadpole}
\ee
Then, canceling out the $m_q$ scale gives
\be
m^2_\sigma \; = \; -8iN_cg^2\int\frac{\dbarfp}{p^2-m_q^2} \; .
\label{msigmas}
\ee
Lastly, the $\sigma$ bubble plus $\sigma$ tadpole graphs of
Fig.~\ref{sigmaloops} in the CL generate the counter-term relation
\begin{vchfigure}[h]
\begin{tabular}{cc}
\includegraphics[scale=0.6]{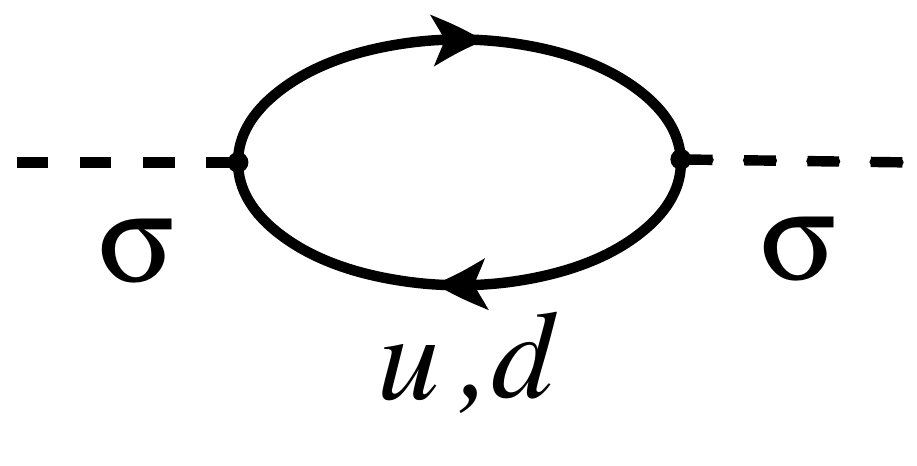} &
\hspace*{15mm}
\includegraphics[scale=0.6]{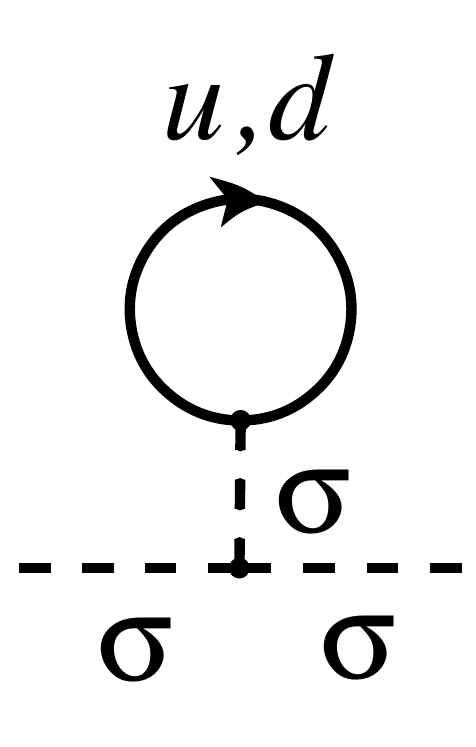} 
\end{tabular}
\mbox{ } \\[-3mm]
\vchcaption{Sigma-meson selfenergy graphs. Left: quark bubble; right:
$\sigma$ tadpole.}
\label{sigmaloops}
\end{vchfigure}
\be
m^2_{\sigma,\cls} \; = \; 16iN_cg^2\!\int\,\dbarfp
\left[\frac{m_q^2}{(p^2-m_q^2)^2}-\frac{1}{p^2-m_q^2}\right]_{\cls} \; .
\label{msigmascl}
\ee
Substituting Eqs.~(\ref{ldge},\ref{msigmas}) into Eq.~(\ref{msigmascl})
leads to the CL relation
\be
m^2_{\sigma,\cls}  \; = \;  -4m^2_{q,\cls}+2m^2_{\sigma,\cls} \;\;\;
\Rightarrow \;\;\; m_{\sigma,\cls}  =  2m_q^{\cls} \; ,
\label{msigmaq}
\ee
but now for the \qllsm\ rather than for the nonlinear NJL model. Note that
Eq.~(\ref{msigmaq}), together with Eq.~(\ref{mconcl}), predicts
$m_\sigma^{\cls}\simeq650.3$~MeV.

Moreover, the integral difference in Eq.~(\ref{msigmascl}) leads to the
dimensional-regula\-rization lemma (DRL) \cite{MPLA10p251:qllsm}, via a Wick
rotation:
\be
\int\dbarfp\left[\frac{m_q^2}{(p^2-m_q^2)^2}-\frac{1}{p^2-m_q^2}\right] \; = \;
-\frac{im^2_q}{16\pi^2} \; .
\label{drl}
\ee
This result follows in many regularization schemes (dimensional, analytic,
$\zeta$-function, Pauli-Villars) and also in a scheme-independent manner
\cite{MPLA13p1893:qllsm,MPLA13p1893_2:qllsm}.
Then, substituting Eq.~(\ref{drl}) back into Eq.~(\ref{msigmascl}) leads to
 \be
m_\sigma^{\cls} \; = \; \frac{\sqrt{N_c}}{\pi}\,g\,m_q^{\cls} \; ,
\label{msigmadrl}
\ee
which predicts, also using Eq.~(\ref{msigmaq}) and $N_c=3$, the crucial
coupling
\be
g \; = \; g_{\pi qq} \; = \; \frac{2\pi}{\sqrt{N_c}} \; \simeq \; 3.6276 \; .
\label{ggpiqq}
\ee
In fact, the latter result also holds
\cite{PRL53p1129:qllsm,PRL53p1129_2:qllsm} in infrared-QCD studies.

Moreover, B.~W.~Lee's null tadpole condition \cite{L72:qllsm}, resulting from
the vanishing of the sum of the three tadpole graphs of Fig.~\ref{null}
\begin{vchfigure}[h]
\begin{tabular}{ccc}
\includegraphics[scale=0.6]{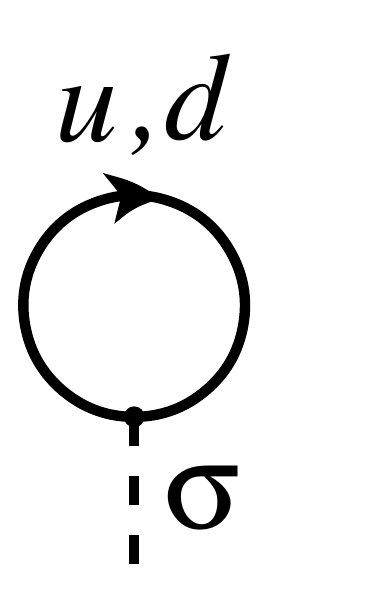} &
\hspace*{12mm}
\includegraphics[scale=0.6]{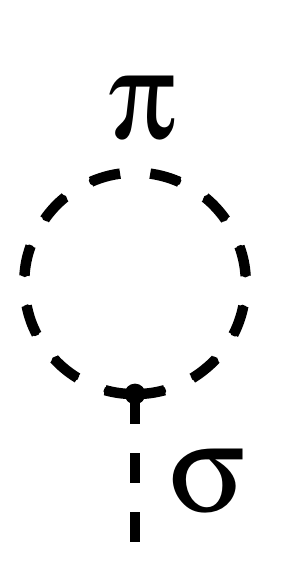} &
\hspace*{14mm}
\includegraphics[scale=0.6]{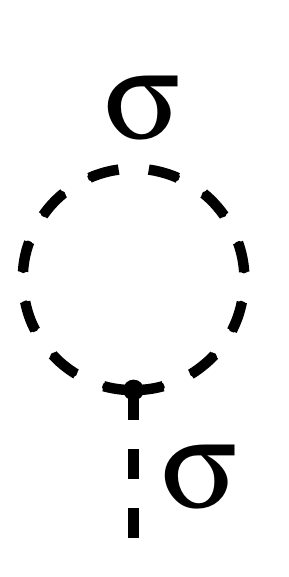} 
\end{tabular}
\mbox{ } \\[-1mm]
\vchcaption{Sigma tadpole graphs. Left: quark loop; middle: $\pi$ loop;
right: $\sigma$ loop.}
\label{null}
\end{vchfigure}
in the CL, reads
\be
0\;=\;\left<\sigma\right>\;=\;-8iN_c\,g\,m_q\int\frac{\dbarfp}{p^2-m_q^2}
\;+\;3ig^2\int\frac{\dbarfp}{p^2}\;+\;3ig'\int\frac{\dbarfp}{p^2-m_\sigma^2}\;,
\label{bwlee}
\ee
where the factors 3 are due to combinatorics.
Now, we drop the middle massless-tadpole term, due to the vanishing of the
$\pi$ mass in the CL, and scale the first and third quadratically divergent
integrals to $m_q^2$ and $m_\sigma^2$, respectively. Using next the identity
$g'=m_\sigma^2/2f_\pi$ along with the GTR $m_q=f_\pi g$, but not needing
Eq.~(\ref{ggpiqq}), the scale $1/f_\pi$ cancels out, which results in
\cite{MPLA10p251:qllsm}
\be
N_c(2m_q^{\cls})^4 \; = \; 3m_{\sigma,\cls}^4 \; .
\label{ncthree}
\ee
Since $m_{\sigma,\cls}=2m_q^{\cls}$ from Eq.~(\ref{msigmaq}), this implies
$N_c=3$. There are indeed many methods to find \cite{MPLA10p251:qllsm} $N_c=3$.

The standard way to verify $N_c=3$ is via the $\pi^0\to2\gamma$ quark-loop
decay amplitude
\be
\left|F_{\pi^0\to2\gamma}\right| \; = \; \frac{\alpha N_c}{3\pi f_\pi}
\simeq 0.02519\;\mbox{GeV}^{-1},\;\;\;\mbox{for}\; N_c=3 \; ,
\label{fpizerotwogamma}
\ee
which predicts a decay rate
\be
\Gamma_{\pi^0\to2\gamma} \; = \;
\frac{m_{\pi^0}^3\left|F_{\pi^0\to2\gamma}\right|^2}{64\pi} \; \simeq \;
7.76\;\mbox{eV} \;.
\label{gpizerotwogamma}
\ee
The latter is very near data \cite{PLB667p1:qllsm}, with
$\tau_{\pi^0}\simeq8.52\times10^{-17}$~s:
\be
\Gamma_{\pi^0\to2\gamma}^{\mbox{\scriptsize PDG}} \; = \;
0.9882\,\frac{\hbar}{\tau_{\pi^0}} \; \simeq \; 7.63\;\mbox{eV} \;,
\label{gpizerotwogammapdg}
\ee
where 0.9882 is the $\pi^0\to2\gamma$ branching fraction.

Since the B.~W.~Lee null tadpole condition (Eq.~(\ref{bwlee})) holds, the
\em true \em \/vacuum corresponds to $\left<\sigma\right>=\left<\pi\right>=0$,
and not to the false vacuum needed in the Gell-Mann--L\'{e}vy
\cite{NC16p705:qllsm,AFFR73:qllsm} nucleon-level \lsm\ for spontaneous
symmetry breaking. Moreover, with $g'=m_\sigma^2/2f_\pi$ in the CL,
the meson-type GTR $g'=\lambda f_\pi$ requires
\be
\lambda \; = \; 2g^2 \;,
\label{lambdag}
\ee
which is valid in both tree and one-loop order, the latter being also true via
the LDGE in Eq.~(\ref{ldge}). This nonperturbative bootstrap scale along with
$g=2\pi/\sqrt{3}$ (Eq.~(\ref{ggpiqq})) requires
\be
\lambda \; = \; 2g^2 \; = \; \frac{8\pi^2}{3} \; \simeq \; 26.3 \; ,
\label{lambda}
\ee
which also holds at one-loop level, owing to the LDGE.

\section{$\bsm{Z=0}$ compositeness condition}
\label{seczezcc}

Following \cite{MPLA10p251:qllsm}, we return to the LDGE integral in 
Eq.~(\ref{ldge}), but now cut off in the ultraviolet (UV) region via a
parameter $\Lambda$. Then, the equation becomes, for $X=\Lambda^2/m_q^2$,
\be
1 \; = \; -i4N_cg^2\int_0^\Lambda\frac{\dbarfp}{(p^2-m_q^2)^2} \; = \;
\ln(X+1)-\frac{X}{X+1} \; .
\label{ldgel}
\ee
This implies $X\simeq5.3$, so that in the CL the UV cutoff scale
becomes
\be
\Lambda \; \simeq \;\sqrt{5.3}\,m_q^{\cls} \;\simeq \;2.302\:
(325.1\;\mbox{MeV}) \;\simeq \;749\;\mbox{MeV} \;.
\label{cutoff}
\ee
Now, the 749 MeV UV scale separates the elementary particles $\pi(137)$
and $\sigma_{\cls}(650)$ from the $q\bar{q}$ bound states $\rho(775)$,
$\omega(783)$, $f_0(980)$, $a_0(980)$, $a_1(1260)$, and so forth. This is
called a $Z=0$ compositeness condition (CC)
\cite{NC25p224:qllsm,NC25p224_2:qllsm,NC25p224_3:qllsm}.  In the
 \qllsm, the condition follows from the renormalization constant being
\be
Z \; = \; 1 - \frac{N_cg^2}{4\pi^2} \; ,
\label{renormalization}
\ee
which vanishes for $N_c=3$, since $g=2\pi/\sqrt{3}$. For more details, we refer
to \cite{NC25p224_3:qllsm}. When meson loops are folded
 in, the UV cutoff equation changes to \cite{JPG24p1:qllsm}
\be
1 \; = \; \ln(X'+1)-\frac{X'}{X'+1} + \frac{1}{6} \; ,
\label{ldgeml}
\ee
where the extra term amounts to $\lambda/16\pi^2$, with $\lambda=8\pi^2/3$
(Eq.~(\ref{lambda})). Given Eq.~(\ref{ldgeml}), we predict $X'\simeq4.15$,
leading to a reduced UV scale
\be
\Lambda^\prime\;\simeq\;\sqrt{4.15}\,m_q^{\cls}\;\simeq\;662\;
\mbox{MeV}\;.
\label{cutoffm}
\ee
This value is quite near $m_{\sigma,\cls}\simeq650$~MeV, which mass even 
increases slightly away from the CL:
\be
m_\sigma \; = \; \sqrt{m_{\sigma,\cls}^2+m_\pi^2} \; \simeq \;
664.1\;\mbox{MeV} \; .
\label{msigmancl}
\ee
Either $\Lambda'\simeq662$~MeV or $m_\sigma\simeq664$~MeV are about
85~MeV less than the usual $Z=0$ CC cutoff at 749~MeV in Eq.~(\ref{cutoff}).
The very similar energy scales in Eqs.~(\ref{cutoffm}) and (\ref{msigmancl}) 
indicate that the inclusion of meson loops leads to a ``double counting'' of
$q\bar{q}$ states as partially elementary and partially bound states. This 
issue is addressed in more detail in \cite{JPG24p1:qllsm}.

Specifically, the nonstrange $q\bar{q}$ pion $\pi(137)$ is an elementary 
particle in the \qllsm, but the also nonstrange $q\bar{q}$ scalar resonance
$\sigma(664)$ can be treated as either elementary or a bound state
\cite{NC25p224_3:qllsm}. This may
be one of the reasons why it has been so difficult to experimentally identify
the $f_0(500)$, with a listed \cite{PLB667p1:qllsm} mass range of
400--550~MeV. Nevertheless, the $\pi$ and the $\sigma$ can be treated in a 
current-algebra fashion as ``chiral partners'' \cite{MPLA17p1673:qllsm}.

\section{Chiral shielding}
\label{secshielding}

The same kind of chiral cancellations as employed in the formulation of
the \qllsm\ in Sects.~\ref{secqllsm} and \ref{secdynamical} can be invoked to
explain the smallness of certain decay or scattering amplitudes, or even their
nonobservation. Here, we shall focus on two processes, viz.\ 
$a_1(1260)\to\pi(\pi\pi)_{\mbox{\scriptsize$S$-wave}}$ and
$\gamma\gamma\to\pi\pi$.

For conserved axial currents ($\partial\cdot A=0$), the leading quark-loop
pion propagator can be shielded via the Dirac matrix \em identity \em
\/\cite{PRD62p037901:qllsm,PRD62p037901_2:qllsm}
\be
\frac{1}{\pslash - m} \, 2m\gamma_5 \, \frac{1}{\pslash - m}\;=\;-\gamma_5 \,
\frac{1}{\pslash - m} - \frac{1}{\pslash - m} \,\gamma_5 \; .
\label{shielding}
\ee
Then, as $p_\pi\to0$, the
$a_1(1260)\to\pi(\pi\pi)_{\mbox{\scriptsize$S$-wave}}$ box and triangle graphs
of Fig.~\ref{aone}
\begin{vchfigure}[h]
\begin{tabular}{cc}
\includegraphics[scale=0.6]{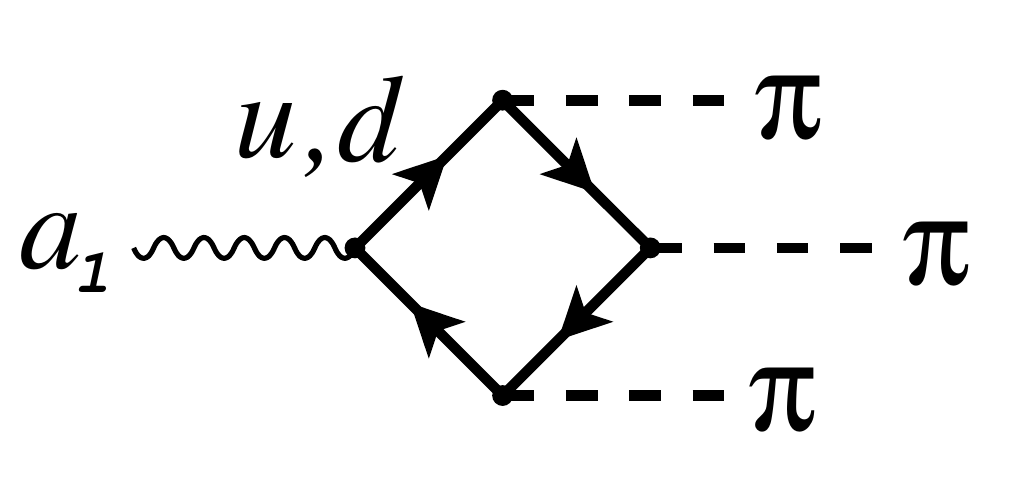} &
\hspace*{3mm}
\includegraphics[scale=0.6]{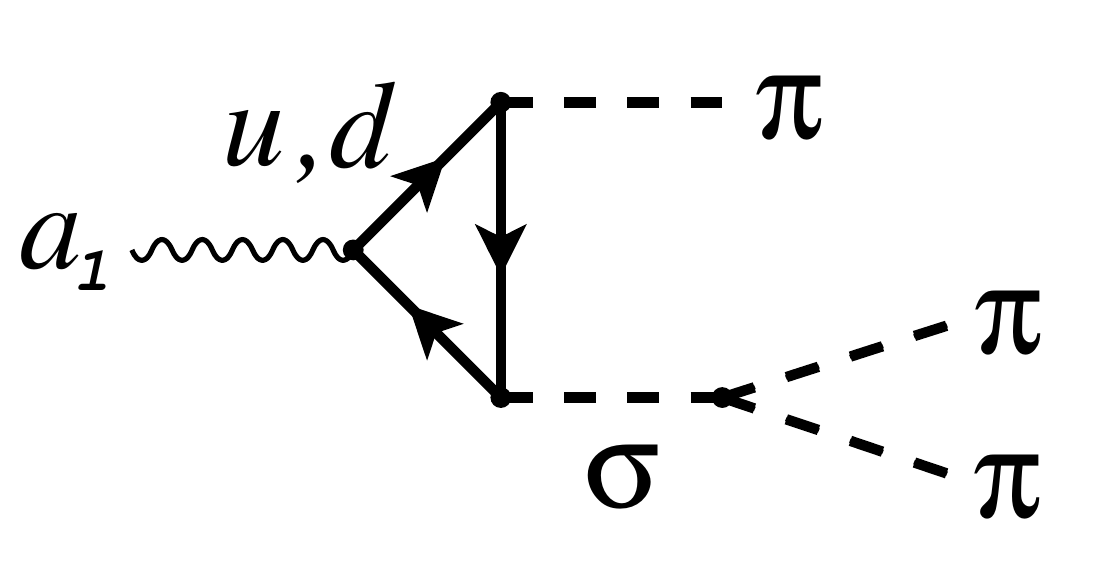} 
\end{tabular}
\mbox{ } \\[-3mm]
\vchcaption{Graphs for the decay \cite{PLB667p1:qllsm} $a_1(1260)\to3\pi$.
Left: quark box; right: quark triangle.}
\label{aone}
\end{vchfigure}
sum up to \em zero, \em in the CL. That is, for $p_\pi\to0$,
\bea
M_{a_1\to3\pi}^{\mbox{\scriptsize box}} & \longrightarrow &
-\frac{1}{f_\pi}M_{a_1\to\sigma\pi} \; , \\
M_{a_1\to3\pi}^{\mbox{\scriptsize triangle}} & \longrightarrow &
\;\;\:\frac{1}{f_\pi}M_{a_1\to\sigma\pi} \; .
\label{aboxtriangle}
\eea
So the total $a_1(1260)\to3\pi$ soft-momentum amplitude is
\be
M_{a_1\to3\pi}^{\mbox{\scriptsize total}} \; = \;
M_{a_1\to3\pi}^{\mbox{\scriptsize box}} +
M_{a_1\to3\pi}^{\mbox{\scriptsize triangle}} 
\; \longrightarrow \; 0 \; ,
\label{atotal}
\ee
which is in agreement with the old experimental decay rate
\be
\Gamma_{a_1(1260)\to\pi(\pi\pi)_{\mbox{\scriptsize$S$-wave}}} \; \lesssim \;
(1\pm1)\;\mbox{MeV} 
\label{apipipi}
\ee
reported in the 1990 PDG tables \cite{PLB239p1:qllsm}, on the basis of the
analysis of \cite{PRD26p82:qllsm}. On the other hand,
the lone $M_{a_1\to\sigma\pi}$ amplitude, which corresponds to only the
triangle graph in Fig.~\ref{aone}, is \em not \em \/small, as confirmed
by the experimental \cite{PRD61p012002:qllsm} decay rate 
\be
\Gamma_{a_1(1260)\to\sigma\pi} \; \sim \; (130\pm40)\;\mbox{MeV} \;.
\label{asigmapi}
\ee
This is one of the cleanest checks of chiral cancellations in the \qllsm.

Another confirmation comes from the process $\gamma\gamma\to\pi\pi$, whose
rate should vanish as $s\to m_\sigma^2$. Namely, just as in the above $a_1$
case, there is a quark-box and a quark-triangle contribution, as depicted
in Fig.~\ref{gammagammapipi}, leading to a total amplitude
\begin{vchfigure}[h]
\begin{tabular}{cc}
\includegraphics[scale=0.6]{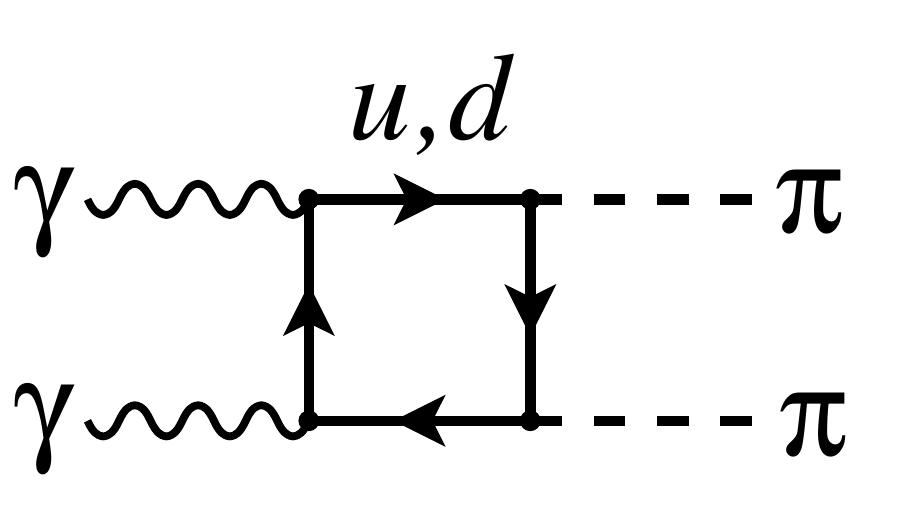} &
\hspace*{5mm} 
\includegraphics[scale=0.6]{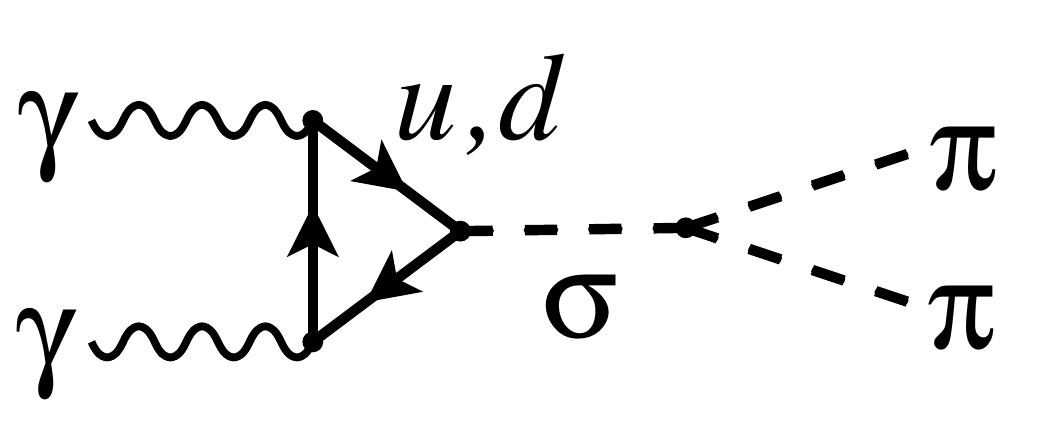} 
\end{tabular}
\mbox{ } \\[-2mm]
\vchcaption{Graphs for the process $\gamma\gamma\to\pi\pi$.
Left: quark box; right: quark triangle.}
\label{gammagammapipi}
\end{vchfigure}
\be
\left<\pi\pi|\gamma\gamma\right>  \; =  \; 
-\frac{\left<\pi\pi|\gamma\gamma\right>_{\mbox{\scriptsize box}}}{f_\pi}+
\frac{\left<\pi\pi\to\sigma|\gamma\gamma\right>_{\mbox{\scriptsize triangle}}}
{f_\pi} \; \rightarrow \;  0 \; .
\label{gammagammapipizero}
\ee
This result is compatible with Crystal Ball data \cite{PRD41p3324:qllsm}, which
revealed a tiny $\gamma\gamma\to\pi^0\pi^0$ cross section of the order of
10~nb at energies around the $\sigma$ mass. Four years earlier, Kaloshin and
Serebryakov \cite{ZPC32p279:qllsm} had predicted a $\gamma\gamma\to\pi^+\pi^-$
cross section exactly of this magnitude at the mass of a then hypothetical
scalar $\epsilon(700)$ resonance. On the other hand, the
$f_0(500)\to\gamma\gamma$ decay rate is quite large, viz.\ of the order of
3--4~keV
\cite{PRL97p011601:qllsm,PRL97p011601_2:qllsm,PLB665p205:qllsm,
PLB665p205:qllsm_2,PRD79p098501:qllsm}.
Again, this is due to the
fact that now only the triangle graph in Fig.~\ref{gammagammapipi} contributes.

As final examples, we should mention the processes $\pi^-p\to\pi^-\pi^+n$
and $K^-p\to K^-\pi^+n$ \cite{PRD62p037901:qllsm,PRD62p037901_2:qllsm}, in
which the $S$-wave amplitudes
for the $\pi^-\pi^+$ and $K^-\pi^+$ final states, respectively, are suppressed
once again because of a cancellation between a box and a triangle diagram.
The triangles correspond to the decays of the scalar resonances $f_0(500)$
(alias $\sigma$) and $K_0^\ast(800)$ (alias $\kappa$), respectively, which
have been so hard to observe experimentally, exactly because of chiral
shielding.

\section{Linking the nonstrange current quark mass scale with the 
$\bsm{\pi N}$ $\bsm{\sigma}$-term}

The PDG \cite{PLB667p1:qllsm} now lists the light current quark masses 
as $m_u=2.3^{+0.7}_{-0.5}$~MeV and $m_d=4.8^{+0.7}_{-0.3}$~MeV, estimated in a
mass-independent subtraction scheme such as $\overline{\mbox{MS}}$, at a scale
$\mu\simeq2$~GeV \cite{PLB667p1:qllsm}. These current quarks are generally
 believed
to be dressed by gluons so as to acquire constituent masses of a few hundreds
of MeV, like $\mhatcon$ in the \qllsm. However, this dressing is a highly
nonperturbative and nonlinear process, which does not allow to write down 
simple relations between current and constituent masses. Nevertheless, 
chiral symmetry does allow to estimate an  {\it effective} \/nonstrange current
quark mass, as the difference between the constituent mass and the dynamical
mass, i.e.,
\be
\mhcur \; = \; \mhatcon -\mdyn \; .
\label{mhatcur}
\ee
Then, in the limit $m_\pi\to0$, we get $\mhcur\to0$ and $\mhatcon\to\mdyn$.
Unfortunately,
Eq.~(\ref{mhatcur}) is only a rough relation, because $\mhatcon$ and $\mdyn$
are on nearby mass shells. To fine-tune Eq.~(\ref{mhatcur}) in the low-energy
region, we invoke infrared QCD, stating that the dynamical
quark mass should run as \cite{EPL80p51001:qllsm}\footnote
{Also see \cite{PRD20p2421:qllsm},
which combined current quark masses with structure functions, finding
$\mhcur\sim62$~MeV and $(m_s/\mhat)_{\scur}\sim5$, close to
$\mhcur\simeq68.3$~MeV and $(m_s/\mhat)_{\scur}\simeq6.26$ above.}
\be
\mdyn(p^2) \; = \; \frac{\mdyn^3}{p^2} \; ,
\label{mdynrun}
\ee
where for consistency
\be
\mdyn \; \equiv \; \mdyn(p^2\!=\!\mdyn^2) \; .
\label{mdynshell}
\ee
On the other hand, $\mdyn$ can be estimated from the nucleon mass as
\be
\mdyn \; \simeq \; \frac{m_N}{3} \; \simeq \; 313\;\mbox{MeV} \; .
\label{mdynn}
\ee
Using now the nonstrange constituent quark mass of $\mhatcon\simeq337.5$
from Eq.~(\ref{nrqm}) above, Eqs.~(\ref{mhatcur}--\ref{mdynn}) yield
\cite{EPL80p51001:qllsm}
\be
\mhcur \; = \; \left(337.5 - \frac{313^3}{337.5^2}\right)\;\mbox{MeV}
\; \simeq \; 68.3\;\mbox{MeV} \; .
\label{mhatcureff}
\ee
Note that this effective current quark mass away from the CL is remarkably
close to half the average pion mass
\be
\bar{m}_\pi \; = \; \frac{139.57+134.98}{2}\;\mbox{MeV} \; \simeq \;
137\;\mbox{MeV} \; .
\label{mpionav}
\ee

Next we study the nucleon, which is a nonstrange $qqq$ state, and
the $\pi N$ $\sigma$ term at the Cheng--Dashen (CD) point
\cite{PRL26p594:qllsm} $\bar{t}\equiv2\bar{m}^2_\pi=2\mu^2$. Early
estimates were \\[5pt] \hspace*{20pt}
$\sigma_{\pi N}(\bar{t})=(66\pm 9)$ MeV \cite{NPB72p310:qllsm}\;,\;\;\; 
$\sigma_{\pi N}(\bar{t})=(65\pm 6)$ MeV \cite{H79:qllsm}\;,\footnote
{Also see \cite{PLB35p445:qllsm}.}  \\[3pt] \hspace*{20pt}
$\sigma_{\pi N}(\bar{t})=(70\pm 6)$ MeV \cite{JPG6p423:qllsm}\;,\;\;\; 
$\sigma_{\pi N}(\bar{t})=(64\pm 8)$ MeV \cite{ZPC15p161:qllsm}\;. \\[5pt]
More recently \cite{PLB482p50:qllsm}, a slightly larger value of $(71\pm9)$~MeV
was found, but still perfectly compatible with the first few analyses. The
average of these five numbers gives $\sigma_{\pi N}(2\mu^2)\simeq67$~MeV,
which is surprisingly close to the effective current quark mass
$\mhcur\simeq68.3$~MeV in Eq.~(\ref{mhatcureff}). Note that both
$\sigma_{\pi N}$ and $\mhcur$ are measures of $SU(2)$ chiral symmetry
breaking.

An overview of the $\pi N$ sigma term can be found in
\cite{PINNL13p362:qllsm}. Starting point is the quenched-lattice
prediction by the APE Collaboration \cite{PLB258p195:qllsm,PLB258p195_2:qllsm}
\be
\sigma_{\pi N,\,\mbox{\scriptsize quenched}}^{\mbox{\scriptsize APE}} \; = \; 
(24.5\pm2.0)\;\mbox{MeV} \; .
\label{ape}
\ee
This result is near the Gell-Mann--Oakes--Renner (GMOR) \cite{PR175p2195:qllsm}
perturbative value
\be
\sigma^{\mbox{\scriptsize GMOR}}_{\pi N} \; = \; \frac{m_{\Xi} + m_\Sigma -
2 m_N}{2} \, \frac{m^2_\pi}{m^2_K-m^2_\pi} 
\simeq 26 \; \mbox{MeV} \; .
\label{gmor}
\ee
The nonperturbative, nonquenched (NQ) addition to $\sigma_{\pi N}$ stems
from the $\sigma$-meson tadpole graphs
\cite{ZPC60p307:qllsm,ZPC60p307_2:qllsm,ZPC60p307_3:qllsm}, yielding
\be
\sigma_{\pi N}^{\mbox{\scriptsize NQ}} \; = \; \frac{m^2_{\pi^0}}{m_\sigma^2}
\,m_N \; = \; \frac{134.98^2}{664.1^2}\,938.9\;\mbox{MeV}  \simeq 
38.8\;\mbox{MeV}\;.\footnotemark
\label{pinnq}
\ee
\footnotetext
{This value is remarkably close to the mass of a hypothetical new light boson,
for which evidence has been found \cite{ARXIV12021739:qllsm} in experimental
data.}
Then, the total $\sigma$ term is predicted to be
\be
\sigma_{\pi N} \; = \; \sigma^{\mbox{\scriptsize GMOR}}_{\pi N} +
\sigma_{\pi N}^{\mbox{\scriptsize NQ}} \simeq (26+39)\;\mbox{MeV}
\; = \; 65\;\mbox{MeV} \; ,
\label{pintotal}
\ee
which is very close to the average value of 67~MeV from the five analyses
above.

The theoretical estimate of 65~MeV is also near the infinite-momentum-frame
(IMF) value \cite{MPLA7p669:qllsm}
\begin{equation}
\sigma_{\pi N} \; = \; \frac{m^2_{\Xi} + m^2_\Sigma - 2 m^2_N}{2 m_N} \,
\frac{m^2_\pi}{m^2_K-m^2_\pi} \; \simeq \; 63 \; \mbox{MeV} \; .
\label{pimimf}
\end{equation}
Note that in the IMF tadpoles are suppressed.

For comparison, the revised chiral-perturbation-theory (ChPT) value now is
60~MeV \cite{HL92:qllsm},\footnote
{Also see \cite{PLB253p252:qllsm,PLB253p252_2:qllsm}.}
which should follow from the positive and coherent sum of \em four \em \/terms,
i.e.,
\begin{eqnarray}
\sigma_{\pi N}(\bar{t}) & = & \sigma_{\pi N}^{\mbox{\scriptsize GMOR}}\;+\;
\sigma_{\pi N}^{\mbox{\scriptsize HOChPT}}\;+\;\sigma_{\pi N}^{\bar{s}s}\;+\;
\sigma_{\pi N}^{t\mbox{-\scriptsize dep.}} \\[1mm]
& \simeq & (25 + 10 + 10 + 15)\;\mbox{MeV} \; = \; 60 \; \mbox{MeV} \; .
\label{sigmachpt}
\end{eqnarray}
Here, the second term on the right-hand side arises from higher-order ChPT,
the third one from the strange-quark sea, and the fourth is a $t$-dependent
contribution due to going from $t=0$ to the CD point,
where the $\pi N$ background is minimal. Leutwyler \cite{HL92:qllsm}
concluded: \em ``The three pieces happen to have the same sign.'' \em Of
course, for things to work out right, all \em four \em \/pieces must have
the same sign, including the GMOR term. Note, however, that
very recently an $N_f\!=\!2$ ChPT analysis of $\pi N$ data
 \cite{1110.3797:qllsm}
managed to extract a value as large as $\sigma_{\pi N}=(59\pm7)$ MeV. The
good news is that, besides $\Delta(1232)$ degrees of freedom,
no contribution from the strange-quark sea was now needed,
in agreement with data \cite{PRL97p102002:qllsm,PRL97p102002_2:qllsm} and
\cite{ZPC60p307:qllsm,ZPC60p307_2:qllsm,ZPC60p307_3:qllsm}, but in stark
conflict with earlier ChPT analyses like in \cite{HL92:qllsm}.  Clearly, the
\qllsm\ amounts to a much simpler and more straightforward approach, 
reproducing the data without any problem.

Summarizing, we have shown is this section that the \qllsm\ effective
current quark mass of $68.3$~MeV is \em very near \em \/the $\pi N$ $\sigma$
term prediction, both via tapoles ($\simeq$ 65~MeV) and using the IMF
($\simeq63$~MeV), which is in turn fully compatible with the experimental
analyses.

\section{Pion charge radius for the L$\bsm{\sigma}$M and VMD schemes}

We now return to \cite{JPG24p1:qllsm}, noting that it took until 1979
\cite{ZPC2p221:qllsm,ZPC2p221_2:qllsm}\footnote{Also see
\cite{PLB205p16:qllsm}.} before the \qllsm\ was employed to calculate the pion
charge radius, in the CL, viz.\
\be
r_{\pi,\cls}^{\qllsms} \; = \; \frac{\hbar c\sqrt{N_c}}{2\pi f_{\pi}^{\cls}} 
\; \simeq \; 0.61\;\mbox{fm} \; ,
\label{pirch}
\ee
for $N_c=3$, $\hbar c\simeq197.3$ MeV$\,$fm, and where $f_{\pi,\cls}$ is now
$\simeq\!89.63$~MeV (see Eq.~(\ref{fpicl})). Stated another way, taking
$g=2\pi/\sqrt{3}$ (as we consistently do) and invoking the GTR
$\mhatcon^{\cls}=f_\pi^{\cls}g\simeq$~325.1 MeV, $r_\pi$ can also be
expressed as
\be
r_{\pi,\cls}^{\qllsms} \; = \; \frac{\hbar c}{\mhatcon^{\cls}} 
\; \simeq \; 0.61\;\mbox{fm} \; .
\label{pirchm}
\ee
Recall that the original vector-meson-dominance (VMD) prediction was
\cite{AOP11p1:qllsm}
\be
r_{\pi,\cls}^{\mbox{\scriptsize VMD}} \; = \; \frac{\sqrt{6}\,\hbar c}{m_\rho}
\; \simeq \; 0.62\;\mbox{fm} \; ,
\label{pirchvmd}
\ee
where we use the PDG \cite{PLB667p1:qllsm} mass $m_\rho\simeq775.5$ MeV.

As for measurements of $r_\pi$, the PDG tables \cite{PLB667p1:qllsm} 
report an average value of $(0.672\pm0.008)$~fm.
There is a tight link between the \qllsm\ and VMD predictions for $r_\pi$,
as stressed in \cite{JPG24p1:qllsm}, viz.\
\be 
\frac{r_{\pi,\cls}^{\mbox{\scriptsize VMD}}}{\hbar c} \; = \;
\frac{\sqrt{6}}{m_\rho} \; \simeq \; \frac{1}{\mhatcon^{\cls}}
\; = \; \frac{\sqrt{3}}{2\pi f_\pi^{\cls}} \; = \;
\frac{r_{\pi,\cls}^{\qllsms}}{\hbar c} \; .
\label{qllsmvmd}
\ee
For quark loops (QL) alone, another (cf.\ Eq.~(\ref{fpiclfpi}))
once-subtracted dispersion relation, evaluated at $q^2=0$, gives in the CL
\cite{NPA724p391:qllsm}
\be
r^2_{\pi,\mbox{\scriptsize QL}} \; = \; \frac{6}{\pi}\int_0^\infty
\frac{\Im\mbox{m}\,F_\pi(q^2)\,dq^2}{(q^2)^2} \; = \;
\frac{N_c(\hbar c)^2}{4\pi^2f^2_{\pi,\cls}} \; ,
\label{rpisql}
\ee
with the form factor normalized to $F_\pi(q^2\!=\!0)=1$. Note that
Eq.~(\ref{rpisql}) precisely amounts to the square of Eq.~(\ref{pirch})
above, which is presumably how
\cite{ZPC2p221:qllsm,ZPC2p221_2:qllsm} arrived at the result.

Concerning the relation between the one-loop-order \qllsm\ with quark loops
alone and the tree-level VMD model, the $Z=0$ compositeness condition and
the cutoff $\Lambda\simeq749$~MeV $<m_\rho$ from Eq.~(\ref{cutoff}) suggest
the $\rho$ meson is an external $q\bar{q}$ bound state. Then, the LDGE in
Eq.~(\ref{ldge}) leads to \cite{MPLA10p251:qllsm,PRD42p941:qllsm}
\be
g_{\rho\pi\pi} \; = \; g_\rho\left[-i4N_cg^2\int^{\Lambda}\!\!
\frac{\dbarfp}{(p^2-m_q^2)^2}\right] \; = \; g_\rho \; ,
\label{grhopipi}
\ee
which is Sakurai's \cite{AOP11p1:qllsm} VMD universality relation.
 
If meson loops (see Fig.~\ref{pionmeson}) are added in Eq.~(\ref{grhopipi}),
the \qllsm\ $g_{\rho\pi\pi}$ coupling  becomes \cite{JPG24p1:qllsm}
\be
g_{\rho\pi\pi} \; = \; g_\rho + \frac{1}{6}\,g_{\rho\pi\pi} \;\;\;
\Longrightarrow \;\;\; \frac{g_{\rho\pi\pi}}{g_\rho}\;=\;\frac{6}{5}\;=\;1.2\;.
\label{grhopipiml}
\ee
Here, $1/6=\lambda/16\pi^2$ as in Eq.~(\ref{ldgeml}).

On the other hand, from data \cite{PLB667p1:qllsm} we get, for $p=363$~MeV and 
$\Gamma_{\rho\pi\pi}=149.1$~MeV,
\be
|g_{\rho\pi\pi}| \; = \; m_\rho\sqrt{\frac{6\pi\Gamma_{\rho\pi\pi}}{p^3}}
\; \simeq \; 5.9444 \; ,
\label{grhopipiexp}
\ee
while the $\rho^0\to e^+e^-$ decay rate $\Gamma_{\rho\bar{e}e}\simeq7.04$~keV
\cite{PLB667p1:qllsm} requires
\be
|g_\rho| \; = \; \alpha\,\sqrt{\frac{4\pi m_\rho}{3\Gamma_{\rho\bar{e}e}}}
\; \simeq \; 4.9569 \; .
\label{grhoee}
\ee
So from data we obtain the ratio
\be
\left|\frac{g_{\rho\pi\pi}}{g_\rho}\right| \; \simeq \; \frac{5.9444}{4.9569}
\; \simeq \; 1.199 \; .
\label{grhopipigrho}
\ee
The agreement with the theoretical \qllsm\ prediction in
Eq.~(\ref{grhopipiml}), which has further improved over the years
\cite{NPA724p391:qllsm}, is simply stunning.

\section{Chiral-symmetry-restoration temperature}
\label{secrestoration}

Next we deal with strong interactions at nonzero temperature, with $\mdyn\to0$
as $T\to T_c$, where $T_c$ is the critical temperature. Now, in the CL
$\mhatcon$ and $m_\sigma$ ``melt'' \cite{PRD31p164:qllsm} to zero at $T_c$,
 according to \cite{IJMPA10p1169:qllsm}
\be
\mhatcon^{\cls}(T) \; = \; \mhatcon^{\cls}-\frac{8N_c\,g^2\,\mhatcon^{\cls}}
{m^2_{\sigma,\cls}}\,\frac{T^2}{2\pi^2}\,\mathcal{J}_+(0)\;,
\label{mhatcontc}
\ee
where 
\be
\mathcal{J}_+(0)\;=\;\int_0^\infty\!\!\frac{x}{e^x+1}\,dx\;=\;
\frac{\pi^2}{12}\;.
\label{jpluszero}
\ee
So at $T=T_c$ the left-hand side of Eq.~(\ref{mhatcontc}) vanishes, yielding
\be
m^2_{\sigma,\cls} \; = \; \frac{N_c\,g^2\,T_c^2}{3} \; .
\label{msigmatc}
\ee
Using $m_{\sigma,\cls}=2\mhatcon^{\cls}$ and the GTR
$\mhatcon^{\cls}=f_\pi^{\cls}g$, we get \cite{PRD39p323:qllsm}, setting
 $N_c=3$,
\be
T_c \; = \; 2f_\pi^{\cls} \; \simeq \; 179.3\;\mbox{MeV}\; .
\label{tcfpi}
\ee

Alternatively, we follow the BCS \cite{PR108p1175:qllsm} procedure, by
first determining the Debye cutoff $k_D$ at $T=0$ \cite{PRD31p164:qllsm}, i.e.,
\be
k_D \; = \; \mdyn\sinh\frac{\pi}{2\alpha_s} \; \simeq \; 1252\;\mbox{MeV}\;,
\label{debye}
\ee
where we have taken $\mdyn\simeq313$~MeV from Eq.~(\ref{mdynn}), and
$\alpha_s\simeq0.75$ at the scale of $m_\sigma$ \cite{PRD31p164:qllsm}.
 So when $\mhatcon(T_c)=0$, or
\be
1 \; = \; \frac{2\alpha_s}{\pi}
\int_0^{\frac{\scriptstyle k_D}{\scriptstyle 2T_c}}\frac{\tanh x}{x}\,dx \; ,
\label{debyeintegral}
\ee
the upper limit of this integral is found to be \cite{PRD31p164:qllsm}
\be
\frac{k_D}{2T_c} \; \; \simeq 3.58 \;\;\; \Longrightarrow \;\;\;
T_c \; \simeq \; \frac{1252}{7.16}\; \mbox{MeV} \; \simeq \;
174.9\;\mbox{MeV} \; .
\label{tcbcs}
\ee

Lastly, one can study the nonlinear NJL \cite{PR122p345:qllsm} model, with
cutoff $\Lambda\simeq749$~MeV (Eq.~(\ref{cutoff})), to derive
\cite{IJMPA10p1169:qllsm}
\be
T_c^2 \; \simeq \; 4f_{\pi,\cls}^2-
\frac{9\hat{m}^4_{\cons,\cls}}{8\pi^2\Lambda^2}\;
\simeq\;(172.8\;\mbox{MeV})^2\;.
\label{tcnjl}
\ee

For comparison, let us just mention\footnote
{See \cite{lattice:qllsm,lattice_2:qllsm,lattice_3:qllsm,lattice_4:qllsm,
           lattice_5:qllsm,lattice_6:qllsm}.}
the results of some lattice computations, which all give a $T_c$ in the
range 157--182~MeV, with error bars accounted for. So our predictions in
Eqs.~(\ref{tcfpi},\ref{tcbcs},\ref{tcnjl}) are fully compatible with the
lattice. Note that the energy scales in these equations
can be converted to a Kelvin temperature scale through
division by the Boltzmann constant $k$.

\section{$\bsm{SU(3)}$ extension of the QLL$\bsm{\sigma}$M}

In order to extend the $SU(2)$ \qllsm\ to $SU(3)$ \cite{IJMPA13p657:qllsm},
we must
first determine the strange constituent quark mass $m_{s,\cons}$. The most
straightforward way to do so, in the context of the \qllsm, is by defining a
GTR for the kaon, viz.\ \cite{JPG5p1621:qllsm,JPG32p735:qllsm}
\be
\frac{m_{s,\cons}+\mhatcon}{2} \; = \; f_K\,g_{Kqq}\;,\;\;\;\mbox{with}\;\;\;
g_{Kqq} \; =\; g_{\pi qq} \; =\; \frac{2\pi}{\sqrt{3}} \; ,
\label{msK}
\ee
where the left-hand side reflects the quark content of the kaon, and
$f_K\simeq1.197f_\pi$ (see \cite{PLB667p1:qllsm}, p.\ 949). Dividing by
$\mhatcon=f_\pi\,g_{\pi qq}$ then yields
\be
\left(\frac{m_{s,\cons}+\mhatcon}{2\mhatcon}\right) \; = \; \frac{f_K}{f_\pi}
\; \simeq \; 1.197 \; ,
\label{fkfpi}
\ee
from which we obtain
\be
\begin{array}{ccl}
m_{s,\cons} \; = \dst\;\left(2\frac{f_K}{f_\pi}-1\right)\,\mhatcon & \simeq & 
1.394\times337.5\;\mbox{MeV} \\[1mm] & \simeq & 470.5\;\mbox{MeV} \; .
\end{array}
\label{mscon}
\ee

We may also estimate $m_s$ roughly from the vector-meson masses $m_\phi$ and
$m_\rho$, since the $\phi(1020)$ is (mostly) $s\bar{s}$ and the $\rho(770)$
is $n\bar{n}$ ($n=u,d$). Using the PDG \cite{PLB667p1:qllsm} masses 1019.5~MeV
and 775.5~MeV, respectively, we thus get
\be
2(m_{s,\cons}-\mhatcon) \; \simeq \; m_\phi-m_\rho
\; \simeq \; 244\;\mbox{MeV} \;,
\label{mphirho}
\ee
which gives $m_{s,\cons}\simeq(337.5+122)$~MeV $\simeq460$~MeV, in reasonable
agreement with the GTR value of 470.5~MeV in Eq.~(\ref{mscon}).

Coming now to the effective strange current quark mass $\mscur$, we may estimate it 
from the kaon and pion masses away from the CL. In a similar fashion as we derived
in Eqs.~(\ref{mhatcureff},\ref{mpionav}) that $\mhcur$ is about half the pion mass,
we now write \cite{EPL80p51001:qllsm} 
\be
\bar{m}_K \; \simeq \; \mscur+\mhcur \; ,
\label{mKmscur}
\ee
where $\bar{m}_K\simeq495.7$~MeV is the average kaon mass
 \cite{PLB667p1:qllsm}.  So this predicts \cite{EPL80p51001:qllsm}
\bea
\mscur \; \simeq \; \bar{m}_K - \mhcur & \simeq & (495.7-68.3)\;\mbox{MeV}
 \nonumber \\ & = & 427.4\;\mbox{MeV} \; .
\label{mscur}
\eea
This in turn gives the ratio \cite{EPL80p51001:qllsm}
\be
\frac{\mscur}{\mhcur} \; \simeq \; \frac{427.4}{68.3} \; \simeq \; 6.26 \; .
\label{mshcur}
\ee
Note that this ratio is much smaller than the value of 25 advocated in ChPT
\cite{PR87p77:qllsm}. However, generalized ChPT \cite{JHEP1p041:qllsm}
 admits quark-mass
ratios considerably smaller than 25. Moreover, also a light-plane approach
\cite{NPB94p163:qllsm} predicts a ratio between 6 and 7, compatible with
Eq.~(\ref{mshcur}).

Let us now use $\mscon$ estimated above to calculate scalar and pseudoscalar
masses besides $m_\sigma$ and $m_\pi$. In
Eqs.~(\ref{mconcl},\ref{msigmaq},\ref{msigmancl}) we determined the isoscalar
scalar mass to be $m_{\sigma,\cls}=2\mhatcon^{\cls}\simeq650.3$~MeV in the
CL, and $m_\sigma=(m_{\sigma,\cls}^2+m_\pi^2)^{1/2}\simeq664.1$~MeV away
from it. Then we predict for the isodoublet scalar $\kappa$ (alias
$K_0^*(800)$ \cite{PLB667p1:qllsm}), using Eqs.~(\ref{mscon},\ref{nrqm}), a
 mass of \cite{EPL80p51001:qllsm}
\be
m_\kappa \; \simeq \; 2\sqrt{\mscon\mhatcon} \; \simeq \; 797\;\mbox{MeV}\;,
\label{mkappa}
\ee
which is very near the E791 data \cite{PRL89p121801:qllsm} at $(797\pm19)$~MeV,
and also compatible with the average of the \em experimental \em
\/masses reported in the PDG listings \cite{PLB667p1:qllsm}.\footnote
{Note that the quoted \cite{PLB667p1:qllsm} \em ``our average'' \em
\/$K_0^*(800)$ mass of $(682\pm29$)~MeV is strongly biased towards the low
value found in a theoretical analysis, and does not represent the average of
the experimental observations.}
Moreover, a mass of 797~MeV is also in reasonable agreement with the $\kappa$
pole positions of $(727-i263)$~MeV, $(714-i228)$~MeV, and $(745-i316)$~MeV
found in the coupled-channel quark-model calculations of
\cite{ZPC30p615:qllsm,ZPC30p615_2:qllsm,ZPC30p615_3:qllsm}, which all
correspond to $S$-wave $K\pi$ resonances peaking at roughly 800~MeV.

Next we shall employ $SU(3)$ equal-mass-splitting laws (EMSLs)
\cite{PRD26p239:qllsm,PRD26p239_2:qllsm,IJMPA13p657:qllsm} to check the
differences between squared scalar and pseudoscalar masses, i.e.,
\be
\begin{array}{l}
m_\sigma^2 - m_\pi^2 \; \simeq \; (0.6641^2-0.1366^2)\;\mbox{GeV}^2
\; \simeq \; 0.42\;\mbox{GeV}^2 \; ,  \\[1mm]
m_\kappa^2 - m_K^2 \; \simeq \; (0.797^2-0.4957^2)\;\mbox{GeV}^2
\; \simeq \; 0.39\;\mbox{GeV}^2 \; , \\[1mm]
m_{a_0}^2 - \bar{m}_{\eta}^2 \; \simeq \;
(0.985^2-0.753^2)\;\mbox{GeV}^2 \; \simeq \; 0.40\;\mbox{GeV}^2 \; ,
\end{array}
\label{esls}
\ee
where $\bar{m}_\eta$ is the average $\eta,\eta'$ mass.
So all three EMSLs have about the same $SU(3)$ chiral-symmetry-breaking scale.

Next we review $\eta$-$\eta'$ mixing. In the flavor
basis $(n\bar{n},s\bar{s})$, $\eta$-$\eta'$ mixing can be written as
\cite{NPB155p409:qllsm,NPB155p409_2:qllsm,NPB155p409_3:qllsm,IJMPA13p657:qllsm}
\bea
\left.|\eta_{n\bar{n}}\right> & = &
\;\;\;\left.|\eta\right>\cos\phi_P+\left.|\eta'\right>\sin\phi_P \\[1mm]
\left.|\eta_{s\bar{s}}\right> & = &
-\left.|\eta\right>\sin\phi_P+\left.|\eta'\right>\cos\phi_P \; ,
\label{etaetap}
\eea
which requires squared masses
\bea
m^2_{\eta_{n\bar{n}}} & = & (m_\eta\cos\phi_P)^2+(m_{\eta'}\sin\phi_P)^2
\label{etansquaredmass} \\[1mm]
m^2_{\eta_{s\bar{s}}} & = & (m_\eta\sin\phi_P)^2+(m_{\eta'}\cos\phi_P)^2 \; ,
\label{etassquaredmass}
\eea
with sum (for any angle)
\be
m^2_{\eta_{n\bar{n}}}+ m^2_{\eta_{s\bar{s}}} \; = \; m^2_{\eta}+ m^2_{\eta'}\;.
\label{metasum}
\ee
From the structure of the pseudoscalar mass matrix, one can then derive 
\cite{NPB155p409:qllsm,NPB155p409_2:qllsm,NPB155p409_3:qllsm,IJMPA13p657:qllsm}
for the mixing angle $\phi_P$ the expressions
\bea
\phi_P & = & 
\arctan\sqrt{\frac{(m^2_{\eta'}-2m^2_K+m^2_\pi)(m^2_\eta-m^2_\pi)}
{(2m^2_K-m^2_\eta-m^2_\pi)(m^2_{\eta'}-m^2_\pi)}} \nonumber \\
       & \simeq & 41.9^\circ
\label{phip}
\eea
or --- equivalently ---
\be
\phi_P \; = \;
\arctan\sqrt{\frac{m^2_{\eta_{n\bar{n}}}-m^2_\eta}
{m^2_{\eta'}-m^2_{\eta_{n\bar{n}}}}} \; \simeq \; 41.9^\circ\;.
\label{phipp}
\ee
In Eq.~(\ref{phip}) we have substituted $m_\eta=547.85$~MeV 
and $m_{\eta'}=957.78$~MeV \cite{PLB667p1:qllsm},
as well as the isospin-averaged kaon and pion masses, while in
Eq.~(\ref{phipp}) the theoretical mass $m_{\eta_{n\bar{n}}}=758.56$~MeV 
from Eq.~(\ref{etansquaredmass}) has been used (Cf.\
$m_{\eta_{s\bar{s}}}=801.29$~MeV from Eq.~(\ref{etassquaredmass})).
This $\phi_P$
is not only well within the wide range $\simeq\!35^\circ$--$45^\circ$ of
experimentally \cite{PLB667p1:qllsm} determined mixing angles, but also close
to the value favored by a coupled-channel model study of the
$a_0(980)\to\pi\eta$ line shape \cite{ZPC30p615_3:qllsm}. Moreover, a mixing
angle of $\phi_P\simeq42^\circ$ allows to reproduce several e.m.\
processes involving the $\eta$ or the $\eta'$, as we shall show in the next
section. Finally, several other works
\cite{PLB234p346:qllsm,PLB234p346_2:qllsm,PLB234p346_3:qllsm,PLB234p346_4:qllsm}
also arrived at a pseudoscalar mixing angle of about $42^\circ$.

To conclude this section, we look at mixing in the scalar-meson sector, viz.\
between the $\sigma$ ($f_0(500)$) and the $f_0(980)$. Now, the mass of the
$f_0(980)$ is known reasonably well, namely at $(990\pm20)$~MeV
\cite{PLB667p1:qllsm}, but the PDG mass of the $\sigma$ is listed
in the wide interval 400--550~MeV, and moreover denoted as ``Breit-Wigner
mass'' or ``$K$-matrix pole'' \cite{PLB667p1:qllsm}. However, the $\sigma$ is
clearly a very broad ($\Gamma\sim$~400--700~MeV) non-Breit-Wigner resonance,
due to the nearby $\pi\pi$ threshold and the Adler zero
\cite{PR137pB1022:qllsm,PR137pB1022_2:qllsm} beneath. So the effective
$\sigma$ mass will always be model dependent, and may very well come out 
above the mentioned range of 400--550~MeV. Then, we may use the
scalar-meson equivalent of Eq.~(\ref{metasum}) to write
\be
m_{\bar{\sigma}} \; = \;
\sqrt{m^2_{\sigma_{n\bar{n}}}+ m^2_{\sigma_{s\bar{s}}}-m^2_{f_0}} \;,
\label{msigmasum}
\ee
where $f_0$ is short for $f_0(980)$.
With the \qllsm/NJL relations $m_{\sigma_{n\bar{n}}}=2\mhatcon\simeq675$~MeV
and $m_{\sigma_{s\bar{s}}}=2\mscon\simeq941$~MeV, we thus estimate
$m_{\bar{\sigma}}\simeq617$~MeV, not too far from $m_\sigma\simeq664.1$~MeV
in Eq.~(\ref{msigmancl}). In order to get the scalar mixing angle $\phi_S$,
we take the scalar versions of
Eqs.~(\ref{etansquaredmass},\ref{etassquaredmass}) and subtract one from
the other, which gives
\be
\begin{array}{ccl}
m^2_{\sigma_{n\bar{n}}}-m^2_{\sigma_{s\bar{s}}} & = &
(m^2_{\bar{\sigma}}-m^2_{f_0})(1-2\sin^2\phi_S) \; = \; \\[2mm]
& & (m^2_{\bar{\sigma}}-m^2_{f_0})\cos2\phi_S \; ,
\end{array}
\label{msigmadif}
\ee
and so
\be
\phi_S \; = \; \frac{1}{2}\arccos\frac
{m^2_{\sigma_{s\bar{s}}}-m^2_{\sigma_{n\bar{n}}}}
{m^2_{f_0}-m^2_{\bar{\sigma}}} \; \simeq \; 21.1^\circ \; .
\label{scalarangle}
\ee
Note that \cite{PLB446p332:qllsm} already estimated $\phi_S\sim20^\circ$
in a similar way.

\section{Electromagnetic decays, quark loops, and meson loops}
\label{strongem}
In this section, we shall compute various e.m.\ decay rates of
pseudoscalar, vector, and scalar mesons using quark loops, and also meson
loops when justified because of phase space.

In terms of a Levi--Civita amplitude $\mathcal{M}$, the rate for a 
pseudoscalar ($P$) meson decaying into two photons is \cite{IJMPA14p4331:qllsm}
\be
\Gamma_{P\to2\gamma}\;=\;\frac{m_P^3|\mathcal{M}_{P\to2\gamma}|^2}{64\pi}\;,
\label{ptwogamma}
\ee
the rate for a vector ($V$) meson decaying into a $P$ meson and a photon
is \cite{IJMPA14p4331:qllsm}
\be
\Gamma_{V\to P\gamma}\;=\;
\frac{p_{P\gamma}^3|\mathcal{M}_{V\to P\gamma}|^2}{12\pi}\;,
\label{vpgamma}
\ee
and the rate for a $P$ meson decaying into a $V$ meson and a photon is
\cite{IJMPA14p4331:qllsm}
\be
\Gamma_{P\to V\gamma}\;=\;
\frac{p_{V\gamma}^3|\mathcal{M}_{P\to V\gamma}|^2}{4\pi}\;.
\label{pvgamma}
\ee
\begin{vchtable}[ht]
\caption{Experimental \cite{PLB667p1:qllsm} and theoretical amplitudes from
 $u$/$d$ quark loops, for several e.m.\ decays of light pseudoscalar and
 vector mesons.  For details on amplitudes, see \cite{IJMPA14p4331:qllsm}.
 Note, however, that the present values of $f_\pi$, $g_\rho$, $\mhatcon$, and
 $\mscon$ have been used.}
\begin{tabular}{||c||c|c|c||}
\hline\hline \mbox{} &&& \\[-3.5mm]
\mbox{Decay} & $\Gamma^{\exps}$ (MeV) & $|\mathcal{M}^{\exps}|$ (GeV$^{-1}$) &
$|\mathcal{M}^{\mbox{\scriptsize th}}|$ (GeV$^{-1}$)\\\hline\mbox{}&&&\\[-3.5mm]
$\pi^0\to\gamma\gamma$ & $(7.74\pm0.55)\times10^{-6}$ & $0.0252\pm0.0009$ &
0.0252 \\
$\eta\to\gamma\gamma$  & $(5.1\pm0.3)\times10^{-4}$ &   $0.025\pm0.001$ &
0.0255 \\
$\eta'\to\gamma\gamma$ & $(4.3\pm0.3)\times10^{-3}$ &   $0.032\pm0.001$ &
0.0344 \\
$\eta'\to\rho\gamma$   & $0.060\pm0.004$ &              $0.41\pm0.02$ &
0.413 \\
$\eta'\to\omega\gamma$ & $(6.2\pm0.5)\times10^{-3}$ &    $0.14\pm0.01$ &
0.152 \\
$\rho^\pm\to\pi^\pm\gamma$ & $0.067\pm0.007$ &          $0.22\pm0.01$ &
0.206 \\
$\rho^0\to\eta\gamma$ & $0.044\pm0.003$ &               $0.48\pm0.02$ &
0.460 \\
$\omega\to\pi^0\gamma$ & $0.70\pm0.03$ &                $0.69\pm0.02$ &
0.617 \\
$\omega\to\eta\gamma$  & $(3.9\pm0.2)\times10^{-3}$ &   $0.14\pm0.01$ &
0.140 \\
$\phi\to\pi^0\gamma$   & $(5.4\pm0.3)\times10^{-3}$ &   $0.040\pm0.002$ &
0.041 \\
$\phi\to\eta\gamma$    & $0.056\pm0.002$ &              $0.21\pm0.01$ &
0.208 \\
$\phi\to\eta'\gamma$   & $2.7\pm0.1$ &                  $0.22\pm0.01$ &
0.212 \\
\hline\hline
\end{tabular}
\label{etaem}
\end{vchtable}
In Eqs.~(\ref{vpgamma},\ref{pvgamma}), $p_{P\gamma}$
and $p_{V\gamma}$ are three-momenta in the decaying
particle's rest frame.

Now we are in a position to analyse several mesonic decays with one or
two photons in the final state. Starting with the two-photon decays of
$P$ mesons, let us recall the famous quark-loop amplitude for
$\pi^0\to\gamma\gamma$, viz.\
\be
|\mathcal{M}_{\pi^0\to\gamma\gamma}| \; = \; \frac{e^2N_c}{12\pi^2f_\pi} \;=\;
\frac{\alpha N_c}{3\pi f_\pi} \; \simeq \; 0.0252 \; \mbox{GeV}^{-1} \; ,
\label{mpizgg}
\ee
where we have substituted $N_c=3$ and \cite{PLB667p1:qllsm} $f_\pi=92.21$~MeV.
The theoretical amplitude is in perfect agreement with
the amplitude extracted from the observed \cite{PLB667p1:qllsm} rate, using
Eq.~(\ref{ptwogamma}),
\be
|\mathcal{M}_{\pi^0\to\gamma\gamma}^{\mbox{\scriptsize exp}}| \; = \;
\left(\frac{64\pi\Gamma_{\pi^0\to\gamma\gamma}}{m_{\pi^0}^3}\right)^{1/2}
\; \simeq \; (0.0252\pm0.0009) \; \mbox{GeV}^{-1} \; .
\label{mpizggexp}
\ee
where we have used the experimental \cite{PLB667p1:qllsm} value
$\Gamma_{\pi^0\to\gamma\gamma}=(7.74\pm0.55)$~eV.
This result encourages us to estimate the pseudoscalar mixing angle 
$\phi_P$ from the observed two-photon widths of the $\eta$ and the
$\eta'$. The amplitudes for $\eta\to\gamma\gamma$ and $\eta'\to\gamma\gamma$
read \cite{IJMPA14p4331:qllsm}
\bea
|\mathcal{M}_{\eta\to\gamma\gamma}|&=&\frac{\alpha N_c }{9\pi f_\pi}
\left(5\cos\phi_P-\sqrt{2}\,\frac{\mhatcon}{m_{s,\cons}}\sin\phi_P\right)\;,
\label{etatwogamma} \\
|\mathcal{M}_{\eta'\to\gamma\gamma}|&=&\frac{\alpha N_c }{9\pi f_\pi}
\left(5\sin\phi_P+\sqrt{2}\,\frac{\mhatcon}{m_{s,\cons}}\cos\phi_P\right)\;,
\label{etaptwogamma} 
\eea
with $\mhatcon\simeq337.5$ MeV and $m_{s,\cons}\simeq470.5$ MeV from
Eqs.~(\ref{mhatfg},\ref{mscon}), respectively. If we now take 
$\phi_P=41.9^\circ$, the latter theoretical amplitudes become
$|\mathcal{M}_{\eta \to\gamma\gamma}|\simeq0.0255$ GeV$^{-1}$ and
$|\mathcal{M}_{\eta'\to\gamma\gamma}|\simeq0.0344$ GeV$^{-1}$, to be
compared with the extracted experimental \cite{PLB667p1:qllsm} ones
$|\mathcal{M}_{\eta \to\gamma\gamma}^{\exps}|\simeq(0.025\pm0.001)$ GeV$^{-1}$
and
$|\mathcal{M}_{\eta'\to\gamma\gamma}^{\exps}|\simeq(0.032\pm0.001)$ GeV$^{-1}$,
respectively. In Table~1, we list the theoretical and experimental
amplitudes of the $\pi^0$, $\eta$, and $\eta'$ $P\to2\gamma$ decays, as
well as those of nine $P\to V\gamma$ and $V\to P\gamma$ processes, several of
which involving an $\eta$ or $\eta'$ meson. For the precise form of the
amplitudes concerning the $P\to V\gamma$ and $V\to P\gamma$ decays, see
\cite{IJMPA14p4331:qllsm}. Let us just mention that, in the case of decays 
involving the $\omega$ or the $\phi$, the small vector mixing angle $\phi_V$,
which expresses the deviation from ideal flavor mixing in this sector, plays
an important role. For instance, the decay $\phi\to\pi^0\gamma$, which would
vanish for ideal mixing, determines to a large extent the value of $\phi_V$,
optimized at $3.8^\circ$ \cite{IJMPA14p4331:qllsm} and allowing to reproduce
the other rates with an $\omega$ or $\phi$ as well. The overall agreement with
data in Table~1 is spectacular, except for the decay $\omega\to\pi^0\gamma$,
which is nevertheless only about 10\% off. 
Finally, the quark-loop approach to e.m.\ decays of mesons, in the
spirit of the \qllsm, also works quite well for several strange, charm,
and even charmonium $V$ states \cite{IJMPA14p4331:qllsm}.

To conclude this section, we consider the two-photon decays of the light scalar
mesons $f_0(500)$ (alias $\sigma$), $f_0(980)$, and $a_0(980)$, with rates
given by
\be
\Gamma_{S\to2\gamma}\;=\;\frac{m_S^3|\mathcal{M}_{S\to2\gamma}|^2}{64\pi}\;,
\label{stwogamma}
\ee
just as in the case of $P$ mesons. Dealing first with the $\sigma$, 
in the NJL limit, i.e., for $m_\sigma=2\mhatcon$, the amplitude takes the
simple form
\cite{PRD69p014010:qllsm}
\be
|\mathcal{M}_{\sigma\to\gamma\gamma}| \; = \; \frac{5\alpha N_c}{9\pi f_\pi}\;,
\label{msigmagg}
\ee
where the factor $5/3$ with respect to the $\pi^0$ amplitude stems from the
fact that, for an isoscalar, the
contributions from the $u$ and the $d$ quark loops add up, in contrast with
the $\pi^0$ case. Assuming that the $\sigma$ is purely
$n\bar{n}=(u\bar{u}+d\bar{d})/\sqrt{2}$, with mass
$m_\sigma^{\mbox{\scriptsize NJL}}=2\mhatcon=675$~MeV, we obtain a
quark-loop rate of 2.70~keV \cite{PRD79p098501:qllsm}.\footnote
{Also see \cite{ARXIV13011567}.}
This rate would becomes 2.57~keV, if we used in Eq.~(\ref{msigmagg}) the value
$m_\sigma=664.1$~MeV from
Eq.~\ref{msigmancl}. But away from the NJL limit, the correct 
gauge-invariant quark-loop amplitude becomes \cite{PRD79p098501:qllsm}
\be
{\cal M}^{n\bar{n}}_{\sigma\to\gamma\gamma} \; = \;
\frac{5\alpha}{3\pi f_\pi} 2\xi_n[2+(1-4\xi_n)I(\xi_n)] \; ,
\label{mnsig}
\ee
where $\alpha=e^2/4\pi$, $\xi_j=m_j^2/m_\sigma^2$,
and $I(\xi)$ is the triangle loop integral given by
\be
I(\xi) \left\{
\begin{array}{ll}
=\;\displaystyle\frac{\pi^2}{2}-2\log^2\left[\sqrt{\frac{1}{4\xi}}+
\sqrt{\frac{1}{4\xi}-1}\:\right] + \displaystyle
2\pi i\log\left[\sqrt{\frac{1}{4\xi}}+\sqrt{\frac{1}{4\xi}-1}\:\right]
\;\;(\xi\leq0.25)\;,\\[5mm]
=\;\displaystyle2\arcsin^2\left[\sqrt{\frac{1}{4\xi}}\:\right]
\;\;(\xi\geq0.25)\;.
\end{array}
\right.
\label{ixi}
\ee
Substitution of $m_\sigma=664.1$~MeV then yields a rate of
2.39~keV, while allowing for an $s\bar{s}$ admixture with scalar mixing angle
$\phi_S=21.1^\circ$ further reduces the rate to 1.84~keV. However, one
now has to include meson loops as well here, which are not negligible at
all, contrary to the $\pi^0$, $\eta$, and $\eta'$
cases, because of phase space. A complete analysis of such contributions was
carried out in \cite{PRD79p098501:qllsm},
including pion, kaon, $\kappa$ ($K_0^*(800)$), and $a_0(980)$ loops. The net
effect of these loops is a very sizable increase of the rate, resulting 
now in a value of 3.39~keV, which should be compared with the recent
analyses yielding 3.1--4.1 keV \cite{PRL97p011601:qllsm,PRL97p011601_2:qllsm}
and 3.1--3.9~keV \cite{PLB665p205:qllsm,PLB665p205:qllsm_2}.
[Note that the quark-loop-only two-photon rate of the $\sigma$ is significantly
smaller than the one reported in \cite{PRD79p098501:qllsm}, due to the
experimentally updated \cite{PLB667p1:qllsm} value $f_K=1.197f_\pi$, leading to
a constituent strange quark mass $m_{s,\cons}=470.5$~MeV, but principally
because of the scalar mixing angle $\phi_s=21.1^\circ$ used here, which is
more realistic than the one employed in \cite{PRD79p098501:qllsm}.
Nevertheless, the
total $\sigma\to2\gamma$ rate, including meson loops, is very close to the
value of $3.5$~keV found in the latter paper. This can be understood from the
interference effects among the various quark-loop and meson-loop
contributions.]

The case of the $f_0(980)$ is trickier, as its mass is
slightly larger than twice the strange quark mass $m_s\simeq470.5$~MeV,
so that we are beyond the NJL limit. Assuming for the moment that this limit
holds approximately, we can estimate the quark-loop amplitude as
\cite{PRD69p014010:qllsm}
\be
|\mathcal{M}_{f_0(980)\to\gamma\gamma}| \; = \;
\frac{\alpha N_cg_{f_0}^{s\bar{s}}}{9\pi\mscon} \; = \;
\label{mfzgg}
\frac{\sqrt{2}\alpha N_c\mhatcon}{9\pi f_\pi\mscon} \; ,
\ee
which gives a two-photon width of about 0.33~keV, compatible with
the average experimental \cite{PLB667p1:qllsm} value
$0.29^{+0.07}_{-0.06}$~keV.
However, much more serious than the small violation of the NJL limit is the 
presence of an $n\bar{n}$ admixture in the $f_0(980)$, corresponding to a
nonvanishing scalar mixing angle, as also suggested by the allowed
\cite{PLB667p1:qllsm} $f_0(980)\to\pi\pi$ decay mode. Namely, the effect of an
$n\bar{n}$ component is enhanced by a factor of roughly 25
 \cite{PRD66p034007:qllsm},
since the electric charge of the $u$ quark is twice that of the $s$ quark,
which makes the prediction of $\Gamma_{f_0(980)\to\gamma\gamma}$ highly
unstable. The necessary inclusion of meson loops, too, will also add to the
uncertainty, although a (partial) cancellation of $n\bar{n}$ and meson-loop
contributions is a plausible possibility. Concretely, including the same meson
loops as above for the $\sigma$, a not unreasonable scalar mixing angle of
$18^\circ$ is required to obtain an $f_0(980)\to2\gamma$ rate of 0.29~keV.
However, caution is recommended because of the very strong
sensitivity of this result to the precise value of $\phi_S$.

Finally, the two-photon width of the $a_0(980)$ is the most difficult one in
the framework of the \qllsm, and probably in any effective description with
quark degrees of freedom. The reason is that the $a_0(980)$ is way beyond the
NJL limit, as  $m_{a_0(980)}=(980\pm20)$~MeV \cite{PLB667p1:qllsm} and
$\mhatcon=337.5$~MeV, so that dispersive effects will arise from the quark
loops. If one simply discards the corresponding imaginary parts --- because of
quark confinement --- and includes meson loops, the \qllsm\ prediction
\cite{PRD69p014010:qllsm} may be compatible with the experimental
\cite{PLB667p1:qllsm} value
$\Gamma_{a_0(980)\to\gamma\gamma}=(0.30\pm0.10)$~keV. 

\section{Scalar mesons $\bsm{a_0(980)}$ and $\bsm{f_0(980)}$}
\label{azfz}

Next we revisit the $a_0(980)$ and $f_0(980)$ scalar mesons, and study
their strong decays.
The PDG tables \cite{PLB667p1:qllsm} now list the isovector $a_0(980)$ and
isoscalar $f_0(980)$ with central masses of $990\pm20$~MeV and
$980\pm20$~MeV, respectively. Henceforth, we shall refer to these scalars
in any equations simply as $a_0$ and $f_0$. In the \qllsm, they are both bound
states heavier than 749~MeV, separated from the elementary $q\bar{q}$ mesons
$\sigma(664)$ and $\pi(137)$, as suggested by the $Z=0$ compositeness
condition in Sec.~\ref{seczezcc} above. Note that also the 
pseudoscalars $\eta_{n\bar{n}}$ and $\eta_{s\bar{s}}$, introduced in the
previous section, are bound states, as
$m_{\eta_{n\bar{n}}}\simeq758.56$~MeV and
$m_{\eta_{s\bar{s}}}\simeq801.29$~MeV.

Now we estimate the strong-interaction decay rate for the process
$a_0\to\eta\pi$, which is approximately given by \cite{PRD69p014010:qllsm}
\be
\Gamma_{a_0\to\eta\pi} \; = \; \frac{p}{8\pi}\left[
\frac{2g_{a_0\eta_{n\bar{n}}\pi}\cos\phi_P}{m_{a_0}}\right]^2
\; \simeq \; 135\;\mbox{MeV} \; ,
\label{azeroetapi}
\ee
using $p=319$~MeV \cite{PLB667p1:qllsm}, $\phi_P\simeq41.9^\circ$, along
with the bound-state CL coupling
\be
g_{a_0\eta_{n\bar{n}}\pi} \; = \;
\frac{m^2_{a_0}-m^2_{\eta_{n\bar{n}}}}{2f_\pi^{\cls}} \; \simeq \;
2.15\;\mbox{GeV} \; ,
\label{gazeroetapi}
\ee
the latter being near the \qllsm\ coupling $\lambda f_\pi^{\cls}=
\left(m^{\cls}_{\sigma}\right)^2\!\!/2f_\pi^{\cls}=
2\mhatcon\,g_{\pi qq}\simeq2.359$ GeV. Furthermore, the PDG tables
report \cite{PLB667p1:qllsm} the branching ratio
$\Gamma_{a_0\to K\bar{K}}/$ $\Gamma_{a_0\to\eta\pi}=0.183\pm0.024$, as well
as the two $a_0\to K\bar{K}$ rates 
$\Gamma_{a_0\to K\bar{K}}\simeq24$~MeV and 25~MeV. On the theoretical side,
the $N/D$ approach to unitarized ChPT \cite{PRD60p074023:qllsm} gives 24~MeV,
while a much earlier analysis \cite{PLB25p294:qllsm} yielded 25~MeV. Thus,
we take the average rate $\Gamma_{a_0\to K\bar{K}}\simeq24.5$~MeV to predict
\be
\Gamma_{a_0\to\eta\pi} \; \simeq \; \frac{\Gamma_{a_0\to K\bar{K}}}{0.183}
\; \simeq \; 134\;\mbox{MeV} \; ,
\label{azeroetapiqllsm}
\ee
which is very near Eq.~(\ref{azeroetapi}) above.

Next we study the $I\!=0\!$ scalar meson $f_0(980)$, and show that it is
mostly an $s\bar{s}$ bound state. Data \cite{PLB667p1:qllsm} finds the e.m.\
branching ratio
\be
\frac{B(\phi(1020)\to f_0\gamma)}{B(\phi(1020)\to a_0\gamma)} \; = \;
\frac{(3.22\pm0.19)\times10^{-4}}{(7.6\pm0.6)\times10^{-5}}
\; = \; 4.24\pm0.42 \; .
\label{phifzazgamma}
\ee
Since we know that the vector $\phi(1020)$ is almost a pure $s\bar{s}$
state, Eq.~(\ref{phifzazgamma}) clearly suggests the isoscalar $f_0(980)$ is
mostly an $s\bar{s}$ scalar bound state (the isovector $a_0(980)$
has no strange-quark content).

Additional information comes from data on strong decays involving the
$f_0(980)$. First of all, there is the non-observation \cite{PLB667p1:qllsm}
 of the
decay $a_1(1260)\to f_0(980)\pi$, whereas $a_1(1260)\to\sigma\pi$ {\em has}
\/been seen \cite{PLB667p1:qllsm}. This again confirms that $f_0(980)$ is
 mainly $s\bar{s}$.  Also, the observed \cite{PLB667p1:qllsm,PAN65p1545:qllsm}
 rate ratio 
\be
\frac{\Gamma_{f_0\to\pi\pi}}{\Gamma_{f_0\to\pi\pi}+\Gamma_{f_0\to K\bar{K}}}
\; = \; 0.84^{+0.02}_{-0.02} \;\;\;\Rightarrow\;\;\;
\frac{\Gamma_{f_0\to K\bar{K}}}{\Gamma_{f_0\to\pi\pi}} \; \simeq \; 0.19 
\label{fzkkpipi}
\ee
is very near the observed branching ratio
$\Gamma_{a_0\to K\bar{K}}/\Gamma_{a_0\to\eta\pi}=0.183$ mentioned above.

Now we estimate the $f_0\to\pi\pi$ partial width as \cite{PRD69p014010:qllsm}
\be
\Gamma_{f_0\to\pi\pi} \; = \; \frac{p}{8\pi}\frac{3}{4}\left[
\frac{2g_{\sigma\pi\pi}\sin\phi_S}{m_{f_0}}\right]^2
\; \simeq \; 42.2 \;\mbox{MeV} \; ,
\label{fzeropipi}
\ee
where we have used $p=471$~MeV \cite{PLB667p1:qllsm}, the \qllsm\ coupling
$g_{\sigma\pi\pi}=\lambda f_\pi^{\cls}=2.357$~GeV, and
$\phi_S\simeq21.1^\circ$ from Eq.~(\ref{scalarangle}).
This partial width is compatible with the E791 \cite{PRL86p765:qllsm} value of
$(44\pm2\pm2)$~MeV, and so lends further support to a scalar mixing
angle of roughly $21^\circ$.

Finally, also weak interactions can be used to show that the $f_0(980)$
is dominantly an $s\bar{s}$ state, by modeling \cite{PLB495p300:qllsm} the
decay $D_s^+\to f_0(980)\pi^+$, with an observed \cite{PLB667p1:qllsm} partial
width of about $2\times10^{-14}$~GeV, via a $W^+$-emission process.

\section{Nonleptonic weak decays $\bsm{K\to2\pi}$}
\label{nonleptonic}
In the present and the next section, we apply our \qllsm\ approach
to nonleptonic kaon decays and the $\Delta I\!=\!1/2$ rule, by introducing 
a weak Hamiltonian.
Nambu \cite{PRL4p380:qllsm} tried to link the (chiral) GTR-conserving axial
currents with semileptonic weak $\pi\to\mu\nu$ decay. Instead, we begin
by extracting the nonleptonic $K\to2\pi$ weak amplitudes from the recent
observed data \cite{PLB667p1:qllsm}, viz.\ (in units of GeV)
\bea
\left|\mathcal{M}_{K_S\to\pi^+\pi^-}\right| \; = \;
\dst m_{K_S}\sqrt{\frac{8\pi\Gamma_{+-}}{q_a}} \; \simeq \;
39.204\times10^{-8}\;, 
\label{kspippim} \\
\left|\mathcal{M}_{K_S\to\pi^0\pi^0}\right| \; = \;
\dst m_{K_S}\sqrt{\frac{16\pi\Gamma_{00}}{q_b}} \; \simeq \;
36.657\times10^{-8}, 
\label{kspizpiz} \\
\left|\mathcal{M}_{K^+\to\pi^+\pi^0}\right| \; = \;
\dst m_{K^+}\sqrt{\frac{8\pi\Gamma_{+0}}{q_c}} \; \simeq \;
1.8125\times10^{-8}\;. 
\label{kppippiz}
\eea
Here, the center-of-mass (CM) momenta (in MeV) $q_a=206$, $q_b=209$, and
$q_c=205$, with decay rates (in $10^{-16}$ GeV) $\Gamma_{+-}=50.825$,
$\Gamma_{00}=22.563$, and $\Gamma_{+0}=0.10995$. The average $\Delta I\!=\!1/2$
scale from Eqs.~(\ref{kspippim},\ref{kspizpiz}) is
\be
\left|\mathcal{M}_{K_S\to2\pi}\right|_{\Delta I=1/2}^{\mbox{\scriptsize avg.}}
\; \simeq \; \frac{39.204+36.657}{2}\times10^{-8}\;\mbox{GeV}
\; \simeq \; 37.93\times10^{-8}\;\mbox{GeV} \; ,
\label{kspipi}
\ee
about 21 times the much smaller $\Delta I\!=\!3/2$ scale in
Eq.~(\ref{kppippiz}).

The first-order-weak (FOW) and second-order-weak (SOW) quark-model-based
scales originate from the SOW soft-kaon theorem \cite{RPP44p213:qllsm},
due to $s$-$d$ single-quark-line (SQL) weak transitions,
generated by the SOW quark loop of Fig.~\ref{sowloop}.
\begin{vchfigure}[h]
\hspace*{10mm}
\includegraphics[scale=0.6]{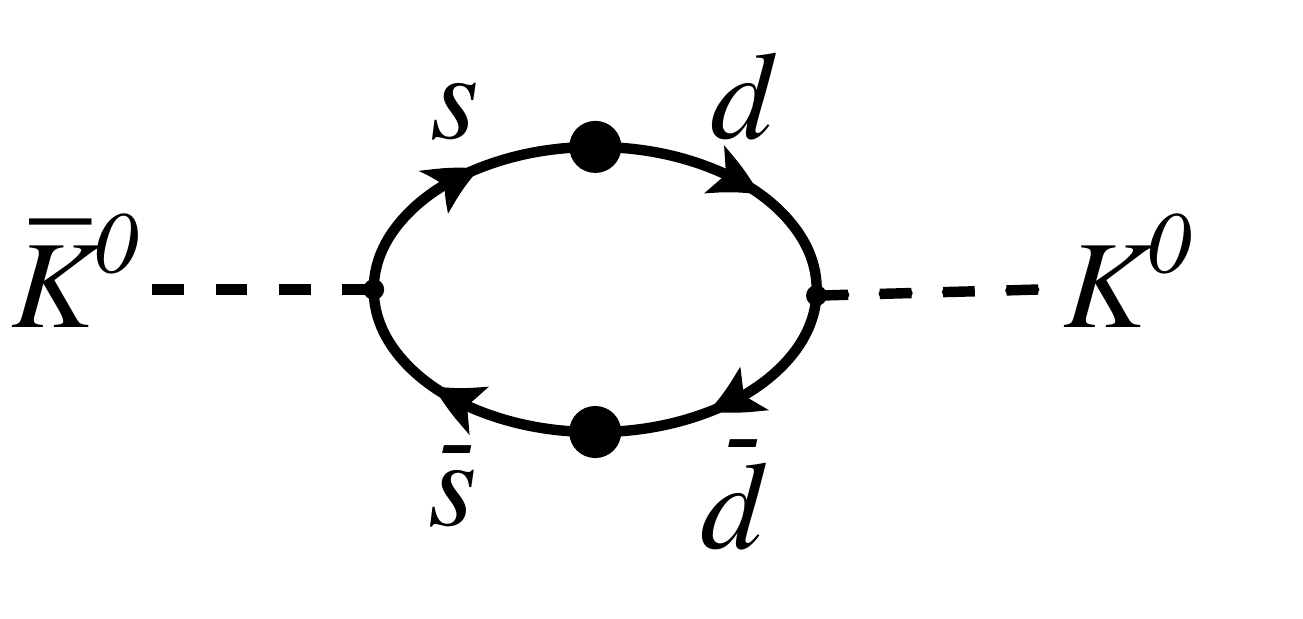} 
\mbox{ } \\[-6mm]
\vchcaption{Second-order-weak $\bar{K}^0\leftrightarrow K^0$ transition.}
\label{sowloop}
\end{vchfigure}
This SOW scale reads
\be
\left|\left<K^0|H_{W,Z}|\bar{K}^0\right>\right| \; = \; 2\beta_w^2m_{K^0}^2
\; = \; m_{K^0}\Delta m_{K_{LS}} \; ,
\label{sqlkzkzb}
\ee
so that the observed \cite{PLB667p1:qllsm}
$\Delta m_{K_{LS}}\simeq3.484\times10^{-12}$~MeV, for \cite{PLB667p1:qllsm}
$m_{K^0}\simeq497.614$~MeV, implies the dimensionless weak scale
\cite{LNC44p193:qllsm}
\be
|\beta_w| \; = \; \sqrt{\frac{\Delta m_{K_{LS}}}{2m_{K^0}}} \; \simeq \;
5.917\times10^{-8} \; .
\label{betaw}
\ee
Note that the GIM scheme \cite{PRL19p1264_3:qllsm} estimated
$|\beta_w|\sim5.6\times10^{-8}$.

Then, the FOW quark loop of Fig.~\ref{fowloop}
\begin{vchfigure}[h]
\hspace*{10mm}
\includegraphics[scale=0.6]{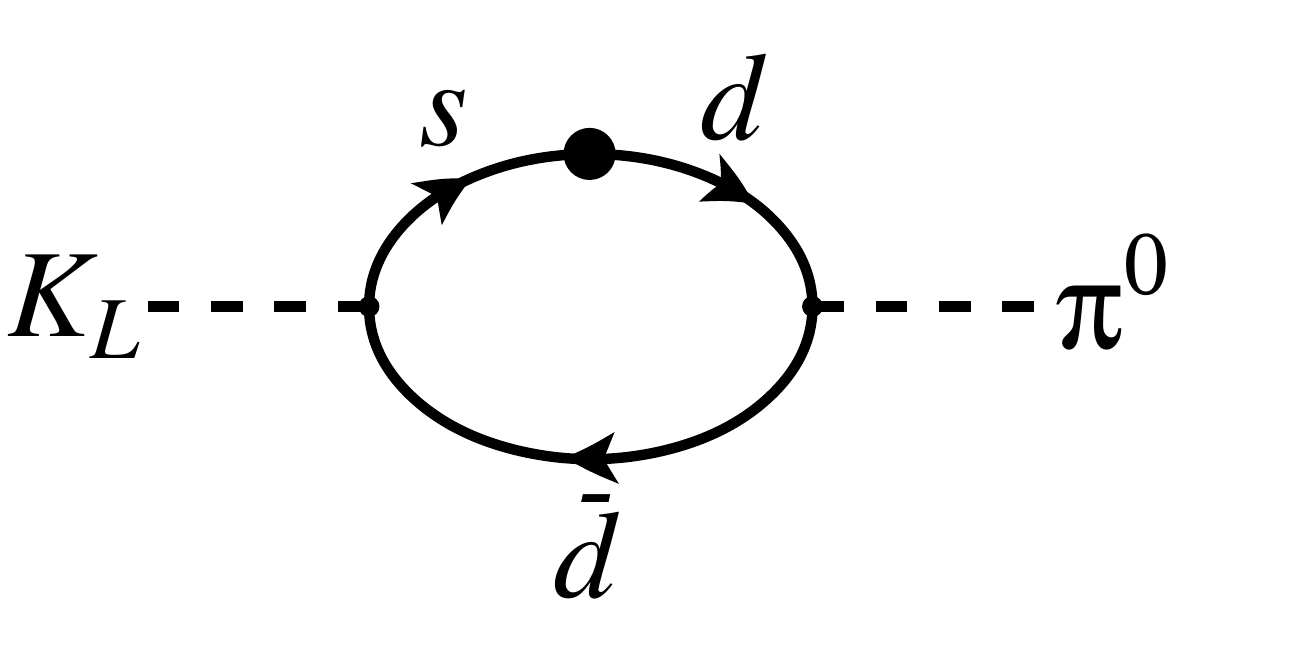} 
\mbox{ } \\[-6mm]
\vchcaption{First-order-weak $K_L\leftrightarrow\pi^0$ transition.}
\label{fowloop}
\end{vchfigure}
generates the $\Delta I\!=\!1/2$ $K_L\to\pi^0$ transition
\cite{LNC44p193:qllsm,MPLA10p1159:qllsm,MPLA19p2267:qllsm}
\be
\left|\left<\pi^0|H_w|K_L\right>\right|_{\Delta I=1/2} \; = \;
\frac{2|\beta_w|m_{K^0}^2f_K}{f_\pi} \; \simeq \;
3.507\times10^{-8}\;\mbox{GeV}^2 \; ,
\label{sqlklpiz}
\ee
with \cite{PLB667p1:qllsm} $f_K/f_\pi\simeq1.197$ and the $\beta_w$ scale from
Eq.~(\ref{betaw}), taking the final-state $\pi^0$ on the $K_L$ mass shell
via PCAC. The crucial FOW scale in Eq.~(\ref{sqlklpiz}) is compatible with
the average of 11 different data sets \cite{MPLA17p2497:qllsm}, such as
 $K\to2\pi$, $K\to3\pi$, $K_S\to2\gamma$, $K_L\to2\gamma$, \ldots:
\be
\left|\left<\pi^+|H_w|K^+\right>\right| \; = \;
\left|\left<\pi^0|H_w|K_L\right>\right| \; = \;
(3.59\pm0.05)\times10^{-8}\;\mbox{GeV}^2 \; .
\label{eleven}
\ee
Note that the predicted $\Delta I\!=\!1/2$ weak scale in
Eq.~(\ref{sqlklpiz}) is only 2.4\% below the central value in
Eq.~(\ref{eleven}). 
Also, the latter equation confirms the scalar
$\sigma$ meson, alias $f_0(500)$ \cite{PLB667p1:qllsm}, as the
``chiral partner'' of $\pi^0$ \cite{MPLA17p1673:qllsm}.

Specifically, we can estimate the $\Delta I\!=\!1/2$ $K\to2\pi$ amplitude via
the $\sigma$-pole graph of Fig.~\ref{sigmapole},
\begin{vchfigure}[h]
\includegraphics[scale=0.6]{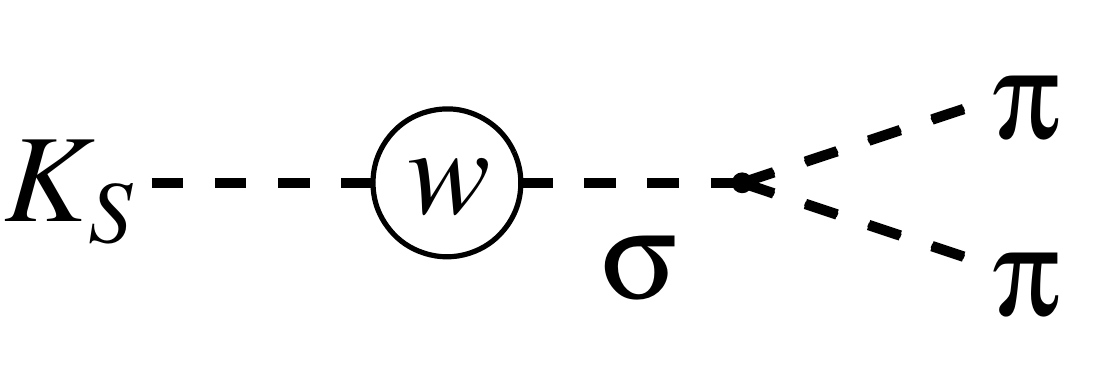}
\mbox{ } \\[-5mm]
\vchcaption{$K_S\to2\pi$ $\sigma$-pole graph.}
\label{sigmapole}
\end{vchfigure}
predicting \cite{MPLA19p2267:qllsm}
\bea
\left|\left<\pi\pi|H_w|K_S\right>\right|_{\Delta I=1/2} & \simeq &
\left|\frac{\left<\sigma|H_w|K_S\right>m_\sigma^2}
{(m_\sigma^2-m_{K^0}^2+im_\sigma\Gamma_\sigma)f_\pi}\right| \\[2mm]
& \simeq & \frac{(3.507\times10^{-8})\:
0.4409}{(0.6654^2)\:0.09221}\;\mbox{GeV}
\; \simeq \; 37.87\times10^{-8}\;\mbox{GeV} \; , 
\label{kspipidih}
\eea
where we have used $m_\sigma\simeq664.1$~MeV, $m_{K^0}=497.6$~MeV, 
$\Gamma_\sigma\simeq600$~MeV, and $f_\pi=92.21$~MeV \cite{PLB667p1:qllsm}.
 Note that
this $\Delta I\!=\!1/2$ estimate lies right between the values from data in
Eqs.~(\ref{kspippim},\ref{kspizpiz}).

Finally, note that pion PCAC (manifest in the nucleon \lsm\
\cite{NC16p705:qllsm,AFFR73:qllsm})
requires, from the weak-interaction chiral commutator $[Q+Q_5,H_w]=0$, that
\bea
\left|\left<\pi\pi|H_w|K_S\right>\right| & \simeq & \displaystyle
\frac{1}{f_\pi}\, \left|\left<\pi|[Q_5^\pi,H_w]|K_S\right>\right| 
\; \simeq \; \displaystyle \frac{1}{f_\pi}\,\left|\left<\pi^0|H_w|K_L\right>
\right| \\[1mm] & \simeq & \displaystyle
\frac{3.507\times10^{-8}\;\mbox{GeV}^2}{0.09221\;\mbox{GeV}} 
\; \simeq \; 38.03\times10^{-8}\;\mbox{GeV} \; .
\label{kshwpipi}
\eea
This scale is again right in between the values in
Eqs.~(\ref{kspippim},\ref{kspizpiz}), and extremely close to the
$\Delta I\!=\!1/2$ estimate in Eq.~(\ref{kspipidih}).

In passing, we confirm that the $\Delta I\!=\!3/2$ scale in
Eq.~(\ref{kppippiz}) is definitely small, as already mentioned above,
by estimating the $W$-emission (WE) $K^+\to\pi^+\pi^0$ amplitude in
Fig.~\ref{wemission},
\begin{vchfigure}[h]
\mbox{ } \\[-4mm]
\hspace*{1mm}
\includegraphics[scale=0.6]{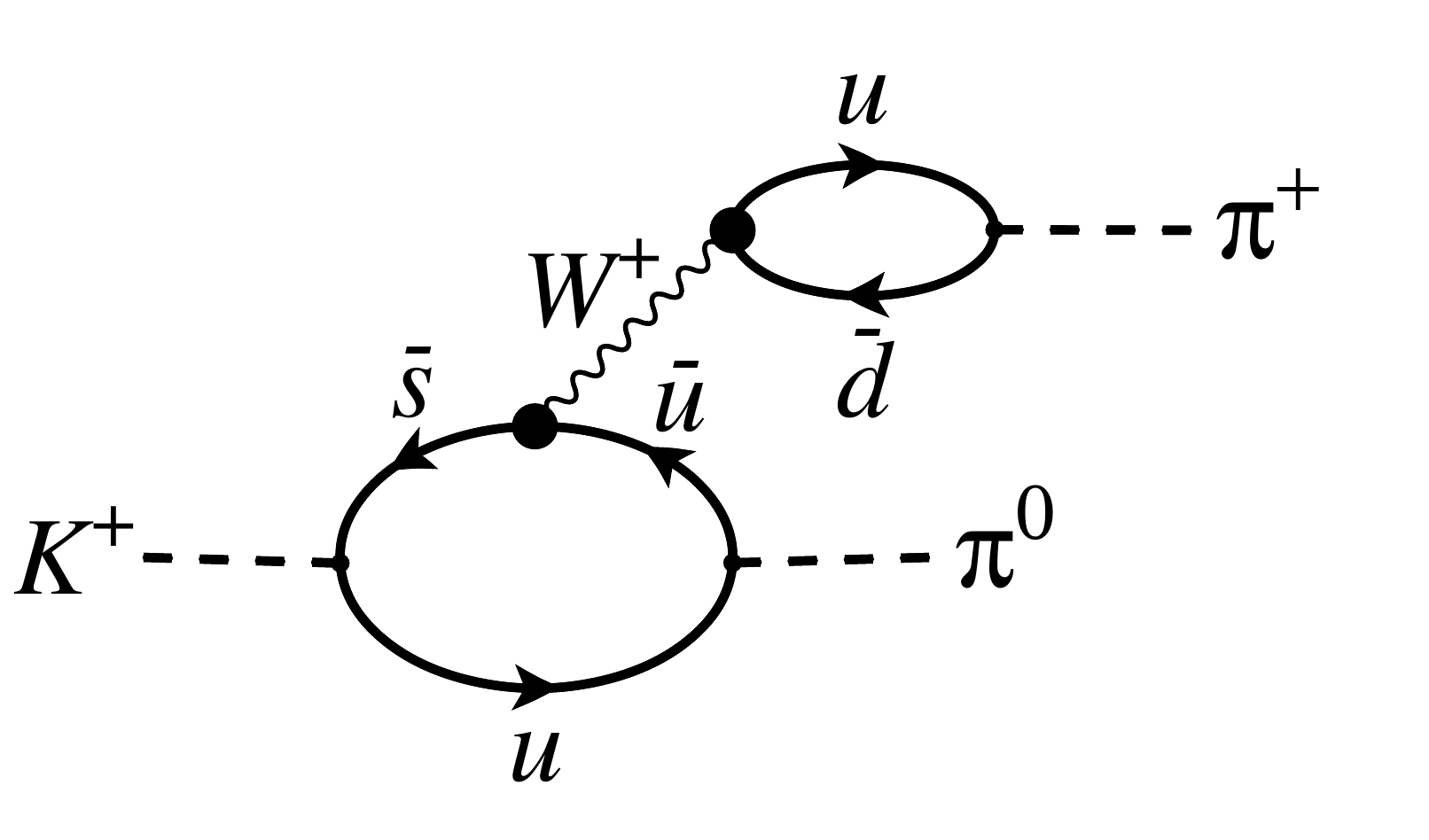} 
\mbox{ } \\[-6mm]
\vchcaption{$W$-emission $K^+\to\pi^+\pi^0$ graph.}
\label{wemission}
\end{vchfigure}
i.e.,
\be
\left|\left<\pi^+\pi^0|H_w|K^+\right>\right|_{\mbox{\scriptsize WE}} \; = \;
\left|\frac{G_FV_{ud}V_{us}}{2\sqrt{2}}\,
(m_{K^+}^2\!-\!m_{\pi^0}^2)f_\pi\right|
\; \simeq \; 1.885\times10^{-8}\;\mbox{GeV} \; ,
\label{we}
\ee
for \cite{PLB667p1:qllsm} $G_F=11.6637\times10^{-6}$~GeV$^{-2}$,
 $|V_{ud}|=0.97419$,
$|V_{us}|=0.2257$, and $f_\pi=92.21$~MeV. This WE estimate is indeed near the
observed $\Delta I\!=\!3/2$ amplitude in Eq.~(\ref{kppippiz}).

\section{$\bsm{K\to2\pi}$ weak tadpole scale and the $\bsm{\Delta I=1/2}$ rule}

As an alternative to estimating the $K_S\to\pi^0\pi^0$ rate via the
$\sigma$-pole graph of Fig.~\ref{sigmapole}, we could compute it through the
tadpole graph of Fig.~\ref{kaontadpole}
\begin{vchfigure}[h]
\includegraphics[scale=0.6]{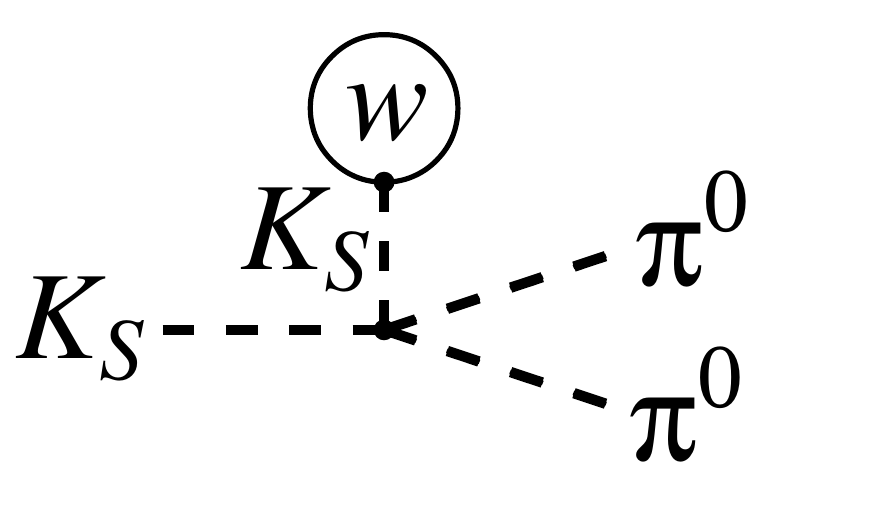}
\mbox{ } \\[-5mm]
\vchcaption{$K_S\to\pi^0\pi^0$ tadpole graph.}
\label{kaontadpole}
\end{vchfigure}
\cite{PLB95p123:qllsm,PLB95p123_2:qllsm,PRD53p2421:qllsm}.
This implies \cite{PRD53p2421:qllsm,MPLA19p2267:qllsm} the amplitude magnitude,
 via PCAC (for $f_\pi=92.21$~MeV),
\be
\left|\mathcal{M}_{K_S\to2\pi^0}\right| \; \simeq \; 
\frac{1}{2f_\pi^2}\left(1-\frac{m_{\pi^0}^2}{m_{K_S}^2}\right)\,
\left|\left<0|H_w|K_S\right>\right| \; ,
\label{mkaontadpole}
\ee
or
\be
\left|\left<0|H_w|K_S\right>\right| \; \simeq \; 
\frac{2f_\pi^2}{1-\displaystyle\frac{m_{\pi^0}^2}{m_{K_S}^2}}\,
\left|\mathcal{M}_{K_S\to2\pi^0}\right| \; \simeq \;
0.6737\times10^{-8}\;\mbox{GeV}^3 \; ,
\label{kshwzero}
\ee
using Eq.~(\ref{kspizpiz}). This scale is not far from the pion PCAC
FOW-SQL weak amplitude, in units of GeV$^3$,
\be
\left|\left<0|H_w|K_S\right>\right| \; \simeq \;
2f_\pi\,2|\beta_w|\,m_{K_S}^2\frac{f_K}{f_\pi} \; \simeq \; 
0.6468\times10^{-8}\;,
\label{kshwzerobeta}
\ee
where we have used $\beta_w\simeq5.9166\times10^{-8}$ from Eq.~(\ref{betaw}),
and \cite{PLB667p1:qllsm} $f_K/f_\pi\simeq1.197$.

Another check on $\left<0|H_w|K_S\right>$ is via the radiative decays
$\pi^0\to2\gamma$ and $K_L\to2\gamma$, the latter process also involving a weak
transition. Given the very successful $\pi^0\to2\gamma$ scale in
Eqs.~(\ref{fpizerotwogamma},\ref{gpizerotwogamma}), for $N_c=3$, the analogue
$K_L\to2\gamma$ amplitude is
\be
\left|F_{K_L\to2\gamma}\right| \; = \;
\sqrt{\frac{64\pi\Gamma_{K_L\to2\gamma}}{m_{K_L}^3}} \; \simeq \;
0.3389\times10^{-8}\;\mbox{GeV}^{-1} \; ,
\label{fkltgamma}
\ee
from the observed rate \cite{PLB667p1:qllsm}
$\Gamma_{K_L\to2\gamma}\simeq0.70376\times10^{-20}$~GeV. Note that,
using the FOW weak scale
$\left<\pi^0|H_w|K_L\right>\simeq3.507\times10^{-8}$~GeV$^{2}$ from
Eq.~(\ref{sqlklpiz}), along with
$\left|F_{\pi^0\to2\gamma}\right|\simeq0.02519$
from Eq.~(\ref{fpizerotwogamma}), the theoretical Levi--Civita
$K_L\to2\gamma$ amplitude obeys
\be
\left|F_{K_L\to2\gamma}\right|_{\mbox{\scriptsize th.}}  \; \simeq \; \dst
\frac{\left|F_{\pi^0\to2\gamma}\right|\left|\left<\pi^0|H_w|K_L\right>\right|}
{m_{K_L}^2-m_{\pi^0}^2} 
\; \simeq \; 0.3851\times10^{-8}\;\mbox{GeV}^{-1}\;.
\label{fkltgammat}
\ee
Then the theoretical radiative tadpole scale is
\be
\left|\left<0|H_w|K_S\right>\right|_{\mbox{\scriptsize th.}}^
{\mbox{\scriptsize rad.}} \; = \; \displaystyle
\left|\frac{F_{K_L\to2\gamma}}{F_{\pi^0\to2\gamma}}\right|\,
(m_{K_L}^2-m_{\pi^0}^2)\,2f_\pi 
\; \simeq \; 0.6468\times10^{-8}\;\mbox{GeV}^3\;,
\label{kshwzerotheory}
\ee
where we have substituted the theoretical $K_L\to2\gamma$ amplitude from
Eq.~(\ref{fkltgammat}). The value in Eq.~(\ref{kshwzerotheory}) is
essentially equal to the pion PCAC FOW-SQL amplitude in
Eq.~(\ref{kshwzerobeta}), which gives us confidence in our tadpole approach.

With hindsight, Weinberg \cite{PRD8p605:qllsm,PRD8p605_2:qllsm}\footnote
{Also see \cite{PRD27p157:qllsm}.}
showed that this ``truly weak''
kaon tadpole \em cannot \em \/be rotated away, as sometimes thought. The
reasonable agreement among the different analyses of the kaon tadpole scale 
in Eqs.~(\ref{kshwzero},\ref{kshwzerobeta},\ref{kshwzerotheory}) confirms
Weinberg's result \cite{PRD8p605:qllsm,PRD8p605_2:qllsm}.

\section{Meson form factors}

Following closely \cite{NPA724p391:qllsm,AIPCP660p311:qllsm}, we study
how meson form
factors, normalized as $F(q^2\!=\!0)=1$, are compatible with the $SU(2)$ and
$SU(3)$ \qllsm\ scheme. Specifically, the charged pion and kaon vector
e.m.\ currents are defined as
\bea
\left<\pi^+(q')|V_{\ems}^\mu|\pi^+(q)\right> & = & F_\pi(q^2)(q'+q)^\mu,
\label{picurrent} \\[1mm]
\left<K^+(q')|V_{\ems}^\mu|K^+(q)\right> & = & F_K(q^2)(q'+q)^\mu .
\label{kcurrent}
\eea
Then the quark loops for the $SU(3)$ \qllsm\ theory predict, in the CL,
\bea
F_{\pi,\qllsms}^{\cls}(k^2) & = & -i4g^2N_c\int_0^1\!\!dx\int\dbarfp 
\left[p^2-\hat{m}_q^2+x(1-x)k^2\right]^{-2} \; , \label{piformfactor} 
\\[1mm]
F_{K,\qllsms}^{\cls}(k^2) & = & -i4g^2N_c\int_0^1\!\!dx\int\dbarfp
\left[p^2-\hat{m}_{sn}^2+x(1-x)k^2\right]^{-2} \; ,
\label{kformfactor}
\eea
where $\hat{m}_q=(m_u+m_d)/2$ and $m_{sn}=(m_s+\hat{m}_q)/2$. The
logarithmic divergence in Eqs.~(\ref{piformfactor},\ref{kformfactor})
can be minimized via a rerouting procedure
\cite{NCA78p159:qllsm,NCA78p159_2:qllsm}, also
using the LDGE in Eq.~(\ref{ldge}), which gives
\bea
F_{\pi,\qllsms}^{\cls}(0) & = & -i4g^2N_c\int\dbarfp
\left[p^2-\hat{m}_q^2\right]^{-2} = 1 \; , 
\label{piformfactorzero} \\[1mm]
F_{K,\qllsms}^{\cls}(0) & = & -i4g^2N_c\int\dbarfp
\left[p^2-\hat{m}_{sn}^2\right]^{-2} = 1 \; , 
\label{kformfactorzero}
\eea
for $m_\pi\to0$ and $m_K\to0$ in the CL. Note the detailed comment in
\cite{NCA78p159:qllsm,NCA78p159_2:qllsm} that rerouting one-half of the
loop momenta in the
opposite direction removes the apparent log divergence of the integrals
in Eqs.~(\ref{piformfactor},\ref{kformfactor}), leading to the finite
integrals in Eqs.~(\ref{piformfactorzero},\ref{kformfactorzero}), 
while also justifying the needed gauge invariance of the vector currents
defined in Eqs.~(\ref{picurrent},\ref{kcurrent}).

Next we study the $\pi^+\to e^+\nu\gamma$ radiative decay form factors, namely
the vector form factor $F_V$ and the axial-vector one $F_A$. Now, the 
experimental status of the latter two observables has varied a lot over
the years. For instance, in 1989 \cite{PLB222p533:qllsm} reported the
measured values $F_V=0.023$ and $F_A=0.021$, albeit with very large error
bars. However, three years earlier the same collaboration measured 
\cite{PLB175p97:qllsm} the ratio $\gamma=F_A/F_V\simeq0.7$, though again with
a huge error. Now, the observed \cite{PLB222p533:qllsm} value $F_V=0.023$ was
in reasonable agreement with the old conserved-vector-current (CVC) prediction
\cite{NC10p342:qllsm}, at a zero value of the invariant $e^+\nu$ mass squared
$q^2$,
\begin{equation}
F_V(0)\;=\;\frac{\sqrt{2}m_{\pi^+}}{8\pi^2f_\pi}\;\simeq\;0.027 \; .
\label{fvcvc}
\end{equation}
On the other hand, the measured \cite{PLB175p97:qllsm} value $\gamma\simeq0.7$
was compatible with the $SU(2)$ \qllsm\ prediction \cite{EPL19p663:qllsm}, from
the sum of a nonstrange quark triangle plus a pion loop,
\be
\gamma \; = \; \frac{F_A(0)}{F_V(0)} \; = \;
1-\frac{1}{3} \; = \; \frac{2}{3} \; .
\label{gammaqllsm}
\ee
However, extension of vector-meson dominance to axial-vector dominance 
\cite{EPL19p663:qllsm} reduces the latter prediction to $\gamma\simeq0.5$.
Much more recently, $F_V$ and $F_A$ have been measured
\cite{PRL103p051802:qllsm} with improved accuracy and an only mild dependence
on $q^2$, resulting in a ratio $\gamma=0.46\pm0.07$, and so compatible with
axial-vector dominance.
 
Next we consider charged-pion polarization for the process
$\gamma\gamma\to\pi^+\pi^-$. The quantities effectively measured are
combinations of the electric and magnetic polarizabilities $\alpha_{\pi^+}$
and $\beta_{\pi^+}$, respectively, viz.\
\cite{PAN56p1595:qllsm,MPLA17p1673:qllsm}
\be
(\alpha-\beta)_{\pi^+} \;\;\; \mbox{and} \;\;\; (\alpha+\beta)_{\pi^+} \; .
\label{polarizabilities}
\ee
Now, general chiral symmetry requires the latter combination to vanish
\cite{PAN56p1595:qllsm}, which is compatible with the experimental result
\cite{ZPC26p495:qllsm}
 $(\alpha+\beta)_{\pi^+}=(1.4\pm3.1_{\mbox{\scriptsize stat}}
\pm2.5_{\mbox{\scriptsize syst}})\times10^{-43}\,\mbox{cm}^3$
($10^{-4}\,\mbox{fm}^3$). So we shall focus on the electric polarizability
$\alpha_{\pi^+}$ only. A simple \qllsm\ estimate predicts 
\be
\alpha_{\pi^+}^{\qllsms}\;\simeq\;
\frac{(\hbar c)^3\alpha_{\mbox{\scriptsize QED}}\,\gamma_{\qllsms}}
{8\pi^2m_\pi f_\pi^2} \; \simeq \; 3.99\times10^{-4}\,\mbox{fm}^3 \; ,
\label{alphaqllsm}
\ee
with $\alpha_{\mbox{\scriptsize QED}}=e^2/4\pi$, $\gamma_{\qllsms}\simeq2/3$
from Eq.~(\ref{gammaqllsm}), and where we have used
$\hbar c\simeq197.3$~MeV$\,$fm, $f_\pi\simeq92.21$~MeV.
A more detailed calculation with quark and pion loops yields
 \cite{PAN56p1595:qllsm}
\be
\alpha_{\pi^+}^{\qllsms}\;\simeq\;
\frac{2.17\,(\hbar c)^3\alpha_{\mbox{\scriptsize QED}}}
{16\pi^2m_\pi f_\pi^2} \; \simeq \; 6.5\times10^{-4}\,\mbox{fm}^3 \; ,
\label{alphaqlpl}
\ee
where the factor $2.17\simeq5/3+0.5$ stems from $N_c$ times the sum of the
squares of the $u$ and $d$ quark charges in the quark loops, i.e.,
$5/3=3\left((2/3)^2+(-1/3)^2\right)$, plus a Feynman-integral
contribution of about 0.5 from the pion loop.
The value of $6.5\times10^{-4}$~fm$^3$ in Eq.~(\ref{alphaqlpl}) agrees
quite well with a recent analysis \cite{PRC73p035210:qllsm} based on several
experiments, resulting in a value 
\be
(\alpha-\beta)_{\pi^+} \; = \; 13^{+2.6}_{-1.9}\times10^{-4}\,\mbox{fm}^3 \; ,
\label{alphaexp}
\ee
which should be divided by two in order to compare with $\alpha_{\pi^+}$, if
one indeed assumes that the sum $(\alpha+\beta)_{\pi^+}$ vanishes or is very
small. These values are also in reasonable agreement with the NJL prediction
\cite{PLB681p147:qllsm} $(\alpha-\beta)_{\pi^+}=9.39\times10^{-4}$~fm$^3$, and
moreover with dispersion sum rules \cite{PLB681p147:qllsm}. However,
\cite{PRL103p051802:qllsm} measured
$\alpha_{\pi^+}=(2.78\pm0.10)\times10^{-4}$~fm$^3$, so that also for this
observable some controversy persists.

For a very recent summary of experimental results on pion
polarizabilites over the years (excluding \cite{PRL103p051802:qllsm}),
see \cite{POSCD09p036:qllsm}, and for a 
detailed discussion of nucleon polarizabilities and their relation to the
two-photon width of the $\sigma$ meson, see
\cite{EPJA30p413:qllsm,EPJA30p413_2:qllsm,EPJA30p413_3:qllsm}.

Lastly, we study the semileptonic weak $K^+\to\pi^0e^+\nu$ ($K_ {\ell3}$)
decay. The nonrenormalization theorem \cite{PRL13p264:qllsm} for the \qllsm\
 says \cite{PRD30p1983:qllsm}
\be
f_+(0)\;=\;1-\frac{g^2}{8\pi^2}\left[\frac{m_{s,\cons}}{\mhatcon}-1\right]^2
\; \simeq \; 0.974 \; ,
\label{fpluszeronr}
\ee
where we have used that $m_{s,\cons}/\mhatcon\simeq2f_K/f_\pi-1\simeq1.394$
\cite{PLB667p1:qllsm}. A prior estimate \cite{AFFR73:qllsm},
based on $K_{\ell2}$ and $K_{\ell3}$ decays, but compatible with the \qllsm,
found
\be
f_+(0)\; =\; \frac{1}{1.23}\frac{f_K}{f_\pi} \; \simeq \; \frac{1.197}{1.23}
\; \simeq \; 0.973 \; .
\label{fpluszeroold}
\ee
Both approaches are in agreement with the data (see
\cite{NPA724p391:qllsm} for details).

\section{Superconductivity and the $\bsm{SU(2)}$ Goldberger-Treiman relation}

Now we shall try to make a link between the energy gap in
superconductivity and the \qllsm, via the critical temperature described in
Sec.~\ref{secrestoration}.

The theory of superconductivity was first understood by Bardeen, Cooper, and
Schrieffer (BCS) in \cite{PR108p1175:qllsm}. In Eq.~(3.30) of this paper,
they found that the ratio of the energy gap $2\Delta$ and the critical
temperature $T_c$ is given by
\be
\frac{2\Delta}{k_BT_c} \; \simeq \; 3.50 \; ,
\label{bcs}
\ee
where $k_B$ is the Boltzmann constant, expressed as
$k_B\simeq8.62\times10^{-5}\:\mbox{eV}\,\mbox{K}^{-1}$
\cite{PLB667p1:qllsm}.
Recall that, at the quark level, with $g_A$=1, and in the CL, the GTR gives,
with $\mdyn\simeq m_N/3\simeq 313$~MeV and $f_\pi^{\cls}\simeq89.63$~MeV,
\be
g_{\pi qq} \; \simeq \; \frac{\mdyn}{f_\pi^{\cls}} \; \simeq \; 3.49 \; .
\label{quarkgap}
\ee
Note that, with $\Delta\to\mdyn$ and $k_BT_c\to2f_\pi^{\cls}$, Eq.~(\ref{bcs})
converts into Eq.~(\ref{quarkgap}). This connection between condensed-matter
and particle physics was stressed in \cite{PB305p175:qllsm}. In some sense,
this is the spirit of Nambu's original work \cite{PRL4p380:qllsm}. Actually,
 the BCS ratio in Eq.~(\ref{bcs}) can be written mathematically as
\cite{PR108p1175:qllsm}
\be
\left(\frac{2\Delta}{k_BT_c}\right)_{\mbox{\scriptsize BCS}} \; = \;
2\pi e^{-\gamma_E} \; \simeq \; 3.528 \; ,
\label{bcseuler}
\ee
where $\gamma_E\simeq0.5772$ is the Euler constant.
On the other hand, the \qllsm\ pion-quark coupling is selfconsistently
bootstrapped to the value (cf.\ Eq.~(\ref{ggpiqq}))
\be
g_{\pi qq} \; = \; \frac{2\pi}{\sqrt{3}} \; \simeq \; 3.628 \; ,
\label{gpiqqbcs}
\ee
still remarkably close to BCS value in Eq.~(\ref{bcseuler}).

Looking directly at condensed-matter phenomenology, data for
2H--$\mbox{NbSe}_2$, with a critical temperature
$T_c=7.2\,\mbox{K}$, finds
\cite{PRL45p660:qllsm,PRL45p660_2:qllsm} an energy gap
$2\Delta=17.2$~cm$^{-1}$, which expressed in the inverse wave length
$1/\lambda=\nu/c=h\nu/hc=E/2\pi\hbar c$ yields
\be
\frac{2\Delta}{k_BT_c} \; \simeq \;
\frac{2\pi\,(197.3\:\mbox{MeV\,fm})\:(17.2\times10^{-13}\:\mbox{fm}^{-1})}
{(8.62\times10^{-5}\:\mbox{eV}\,\mbox{K}^{-1})\:(7.2\,\mbox{K})} 
\; \simeq \; 3.44 \;.
\label{expone}
\ee
Another experiment \cite{PRL70p3987:qllsm}, using an $\mbox{Rb}_3\mbox{C}_{60}$
superconductor, reports a ratio $\Delta/k_B=53\,\mbox{K}$ for
$T_c=29.4\,\mbox{K}$, which gives
\be
\frac{2\Delta}{k_BT_c} \; = \; \frac{106}{29.4} \; \simeq \; 3.61 \;.
\label{exptwo}
\ee
So the average of these two experimental results is very close to the
theoretical BCS predictions in Eqs.\ (\ref{bcs},\ref{bcseuler}), but remarkably
enough also to the \qllsm\ values in Eqs.\ (\ref{quarkgap},\ref{gpiqqbcs}).
As a matter of fact, Nambu found \cite{PR117p648:qllsm}, not long after the
pioneering BCS paper \cite{PR108p1175:qllsm}, that gauge invariance is a valid
concept for superconductivity, but with radiative photons replaced by
acoustical phonons.

\section{Dynamically generating the top-quark and scalar-Higgs masses}
\label{tophiggs}

In this section we shall apply \qllsm\ ideas to the gauge bosons
$W^\pm$ and $Z$, as well as the top quark and Higgs boson, by analogy with
the low-energy sector.

The Higgs mass in the electroweak Standard Model (EWSM) is a free parameter,
but experiment now indicates a direct lower-mass search limit of about 114~GeV,
with 95\% CL \cite{PLB667p1:qllsm}. On the other hand, a global fit to
precision electroweak data, gathered in the course of many years at LEP,
Tevatron, and other accelerators yields a range of $(m_H=94^{+29}_{-24})$~GeV
($m_H<152$~GeV, 95\% C.L.) \cite{PLB667p1:qllsm}. However, these estimates are
based on \em perturbative \em \/EWSM calculations, which leaves some room for
alternative scenarios, also due to the triviality problem
\cite{PLB391p144:qllsm,PLB301p203:qllsm}.\footnote
{Also see the references in \cite{PLB391p144:qllsm} to earlier work on
$\lambda\phi^4$ theory.}
Moreover, the analyses themselves may be less trustworthy than generally
assumed \cite{MPLA10p845:qllsm,MPLA10p845_2:qllsm,MPLA10p845_3:qllsm}.
Finally, clear Higgs-like signals have been seen very recently by the
ATLAS \cite{PLB716p1:qllsm} and CMS \cite{PLB716p30:qllsm} Collaborations
at LHC, i.e., at a mass of about 125~GeV. Further experiments will be needed,
though, to confirm and determine the quantum numbers of the observed boson.

Here, we shall try to estimate the Higgs mass in a \em nonperturbative \em
\/(NP) framework, in the spirit of the \qllsm.

The EWSM couplings are
$g_W^2/8m_W^2=G_F/\sqrt{2}$, for \cite{PLB667p1:qllsm}
$G_F=11.6637\times10^{-6}$~GeV$^{-2}$, which gives, for $m_W=80.385$~GeV,
\cite{PLB667p1:qllsm}
\be
|g_w| \; = \; \sqrt{\frac{8G_F}{\sqrt{2}}}\,m_W \; \simeq \; 0.65295 \; .
\label{gw}
\ee
Furthermore, the NP vacuum expectation value (VEV) $f_w$ is extracted from
$G_F$ as
\be 
f_w \; = \frac{1}{\sqrt{\sqrt{2}G_F}} \; \simeq \; 246.22 \; \mbox{GeV} \; .
\label{fw}
\ee
Then, quadratically divergent tadpole graphs can be made to cancel, in the
spirit of B.~W.~Lee's NP null tadpole condition \cite{L72:qllsm}
(see Fig.~\ref{null}), which allows to predict
\cite{JPG32p735:qllsm,APPB12p437:qllsm,APPB12p437_2:qllsm,APPB12p437_3:qllsm,APPB12p437_4:qllsm}
the scalar Higgs mass. Alternatively, and in analogy with the scalar $\sigma$
meson (cf.\ Eqs.~(\ref{mhat},\ref{msigmaq})), the EWSM Higgs boson is 
dominated by a scalar $t\bar{t}$ state, which in the CL would yield a mass
$m_H^{\cls}=2m_t$. Beyond the CL this value is then reduced to
\be
m_H \; = \; \sqrt{(2m_t)^2-(2m_W^2+m_Z^2)} \; \simeq \; 314.9 \; \mbox{GeV} \;,
\label{mhiggsewsm}
\ee
for the measured \cite{PLB667p1:qllsm} heavy masses (in GeV)
\be
m_t \; \simeq \; 173.5 \;, \;\;\; m_W \; \simeq \; 80.385 \; ,
\;\;\; m_Z \; \simeq \; 91.1876 \; ,
\label{mtwz}
\ee
where we have dropped the (small) errors as well as the negligible
contributions from the much lighter other quarks.

Alternatively, we can use the modified
Kawarabayashi-Suzuki-Riazuddin-Fayyazudin (KSRF) \cite{PRL16p255:qllsm}
 approach in
the electroweak sector. As an illustration, for the $\rho$ meson the KSRF
relation predicts
\be
m_\rho \; = \; \sqrt{2}\,g_{\rho\pi\pi}f_\pi \; \simeq \; 775.18\; \mbox{MeV}\;,
\label{rhoksrf}
\ee
which is remarkably close to the PDG \cite{PLB667p1:qllsm} value
 $m_\rho=775.49$~MeV.
The weak-interaction KSRF analogue is obtained by the substitutions
\cite{JPG32p735:qllsm}
\be
m_\rho\to M_W \;\;\; , \;\;\; g_{\rho\pi\pi}\to\frac{g_w}{2}
\;\;\; , \;\;\; \sqrt{2}\,f_\pi\to f_w \; ,
\label{ksrfw}
\ee
where the weak coupling simulates $g_\rho\tau^+/2$, and the charged $W$
requires a $\sqrt{2}$ in the weak (VEV) decay constant. Thus, the
strong-interaction KSRF relation in Eq.~(\ref{rhoksrf}) translates into
the NP weak KSRF relation
\be
m_W \; = \; \frac{1}{2}\,g_w\,f_w \; ,
\label{mw}
\ee
which is precisely the famous EWSM relation from Eqs.~(\ref{gw},\ref{fw})
\cite{PRL19p1264:qllsm,PRL19p1264_2:qllsm,PRL19p1264_3:qllsm}.
Moreover, using $\Gamma_{Z\to e^+e^-}=83.91$~MeV \cite{PLB667p1:qllsm}, one
 can predict \cite{JPG32p735:qllsm}, via VMD,
\be
|g_Z| \; = \; e\bar{e}\sqrt{\frac{m_Z}{12\pi\Gamma_{Z\to e^+e^-}}}
\; \simeq \; 0.5076 \; ,
\label{gz}
\ee
as $e^2/4\pi\simeq1/137.036$ and $\bar{e}^2/4\pi\simeq1/128.93$ (at the
$m_Z$ scale).

Now, the tree-level vector and axial-vector couplings of the $Z$ to the
electron get modified, from $g^e_V = -1/2 + 2 \sin^2\theta_w$ and
$g^e_A = -1/2$ to \cite{PLB667p1:qllsm}
\be
g^e_A \; = \; -0.5064(1) \;\;\; , \;\;\; g^e_V \; = \; -0.0398(3) \; ,
\label{geavexp}
\ee
by radiative corrections. Nevertheless, $Z$ remains largely axial, since
$\sin^2\theta_w = 0.23108(5)$ (see \cite{PLB667p1:qllsm}, review on
electroweak model, Table 10.2, NOV scheme). The difference $V\!-\!A$ coupling
becomes
\be
g^e_{V\!-\!A} \; = \; 0.4666 \; ,
\label{gevmaexp}
\ee
which is reasonably close to the EWSM value $2\sin^2\Theta_w = 0.462$.
This is also supported by the ratio of Eq.~(\ref{mw}) to the conventional
EWSM rate for the $Z$, namely
\be
\Gamma_{Ze^+e^-}\; =\; \left(\frac{g_w}{4}\right)^2 \frac{M_Z^3}{12\pi M_W^2}
\; = \; 82.94\;\mbox{MeV} \; ,
\label{gammazepem}
\ee
which is near the experimental value of 83.91~MeV.
This yields the alternative NP expression
\be
\sin^2\Theta_w \; = \; 1 - (g_wg_Z/4e\bar{e})^2 \; = \; 0.2316 \;,
\label{sinthetaw}
\ee
and so $2\sin^2\Theta_w=0.4632$, again close to the value of $g^e_{V\!-\!A}$
in Eq.~(\ref{gevmaexp}).

We may now estimate the very heavy top-quark mass $m_t$ via a NP GTR, as we
did for the lighter quarks. Here we have to be careful to
take account of an EWSM factor of $2\sqrt{2}$ and the (V-A) VMD coupling
$g_Z/2$. In this way we get \cite{JPG32p735:qllsm}
\be
m_t \; = \; 2\sqrt{2}\,f_w\,\frac{|g_Z|}{2} \; = \; \sqrt{2}\times
246.22\:\mbox{GeV}\times0.5076 \; \simeq \; 176.8\;\mbox{GeV} \; ,
\label{mt}
\ee
not far away from the PDG \cite{PLB667p1:qllsm} value of
 $(173.5\pm0.6\pm0.8)$~GeV.
If we now use our theoretical prediction for $m_t$ in Eq.~(\ref{mt}) to
estimate the Higgs mass via Eq.~(\ref{mhiggsewsm}), we find about
322~GeV. It is curious to notice that, very recently, indications of a narrow
325~GeV scalar resonance were found \cite{PLB718p943:qllsm} in unpublished,
``exotic'' CDF data \cite{CDF10603:qllsm}.\footnote
{Also see \cite{PRL108p111804:qllsm}.}

Lastly, we remark that the top-quark width is huge, viz.\
$2.0^{+0.7}_{-0.6}$~GeV \cite{PLB667p1:qllsm}, strongly dominated by the
$t\to bW^+$ mode. Note that the latter process is purely weak, though not at
all weak in the common sense. Therefore, $t\bar{t}$ physics, and in particular
a scalar $t\bar{t}$ state, should be dealt with by NP methods. Thus, we think
the Higgs might be selfconsistently described as such a state, being both
elementary and composite, just like the $\sigma$ meson in the \qllsm.

In the next section, we shall revisit the Higgs mass in the context of CP
violation.

\section{CP violation and Higgs mass}

In this section, we deal with CP violation (CPV), but in an NP approach
{\it within} \/the Standard Model, inspired by the \qllsm, and based on a
careful analysis of the most recent CPV data. Moreover, we model CPV in a way
that allows to make another estimate of the scalar Higgs mass.

In \cite{PRD53p2421:qllsm,MPLA19p2267:qllsm}, it was noted that the
presently
observed \cite{PLB667p1:qllsm} branching ratio (BR) for CP-conserving (CPC)
$K_S\to2\pi$ weak decays is extremely near the CPV $K_L\to2\pi$ BR, i.e.,
\bea
B\left(K_S\,\to\,\frac{\pi^+\pi^-}{\pi^0\pi^0}\right)_
{\!\mbox{\scriptsize CPC}} & = &
\displaystyle\frac{(69.20\pm0.05)\,\%}{(30.69\pm0.05)\,\%} \; = \;
2.255\pm0.004 \; , \label{bcpc} 
\\[3mm]
B\left(K_L\,\to\,\frac{\pi^+\pi^-}{\pi^0\pi^0}\right)_
{\!\mbox{\scriptsize CPV}} & = &
\displaystyle\frac{(1.967\pm0.010)\times10^{-3}}
{(8.64\pm0.06)\times10^{-4}} \; = \; 2.277\pm0.02 \; .
\label{bcpv} 
\eea
Not only are the CPC and CPV scales in Eqs.~(\ref{bcpc},\ref{bcpv}) near
each other, but the CPC scale in Eq.~(\ref{bcpc}) is near
 \cite{MPLA19p2267:qllsm}
the $\Delta I\!=\!1/2$ scale of 2, due to the $\sigma$ pole and also the
tadpole graph for $K_S\to2\pi$ weak decays. 
Also, the present CPC radiative BR \cite{PLB667p1:qllsm} is about
\be
B\!\left(\!K_S\,\to\,\frac{\pi^+\pi^-\gamma}{\pi^+\pi^-}\right)_
{\!\mbox{\scriptsize CPC}} \; \simeq \; \displaystyle
\frac{1.79\times10^{-3}}{69.2\times10^{-2}}
\; \simeq \; 2.59\times10^{-3} \; .
\label{bksgamma}
\ee
Moreover, the approximate CPV radiative BR \cite{PLB667p1:qllsm} is
\be
B\!\left(\!K_L\to\frac{\pi^+\pi^-\gamma}{\pi^+\pi^-}\right)_
{\!\mbox{\scriptsize CPV}} \, \simeq \, \dst
\left|\frac{\eta_{+-\gamma}}{\eta_{+-}}\right|^2
B\!\left(\!K_S\to\frac{\pi^+\pi^-\gamma}{\pi^+\pi^-}\right)_
{\!\mbox{\scriptsize CPC}} \, \simeq \, (2.87\pm0.16)\times10^{-3} \; , 
\label{bklgamma} 
\ee
using \cite{PLB667p1:qllsm} $|\eta_{+-\gamma}|=(2.35\pm0.07)\times10^{-3}$ and
$|\eta_{+-}|=(2.232\pm0.011)\times10^{-3}$.
Equation~(\ref{bklgamma}) is 11\% higher than 
Eq.~(\ref{bksgamma})
\cite{MPLA19p2267:qllsm,PRD53p2421:qllsm,IJMPA11p271:qllsm}, whereas
the radiative e.m.\ scale is
\be
\frac{\alpha}{\pi} \; \simeq \; \frac{1}{137.036\,\pi} \; \simeq \;
2.323\times10^{-3} \; ,
\label{alphapi}
\ee
which is 3.9\% higher than $|\eta_{+-}|$ above. Now, it was remarked long
ago \cite{PR139pB1650:qllsm,PR139pB1650_2:qllsm} that the scale of
$\alpha/\pi$ could be the relevant measure for CPV $K\to2\pi$ weak decays. 

In the spirit of the present \qllsm\ review, we follow
\cite{MPLA19p2267:qllsm}
and theoretically compute the radiatively (rad) corrected $\Delta I\!=\!1/2$
SQL scale for $K^0\to\pi^+\pi^-$ minus unity, i.e.,
\be
\left|\frac{\eta_{+-}}{\eta_{00}}\right|_
{\stackrel{\mbox{\scriptsize rad}}{\mbox{\scriptsize SQL}}}\:-\:1 \; = \;
2\,\frac{\alpha}{\pi} \; = \; 4.646\times10^{-3} \; ,
\label{etatheory}
\ee
near CPV data minus unity, viz.\
\be
\left|\frac{\eta_{+-}}{\eta_{00}}\right|_
{\stackrel{\mbox{\scriptsize CPV}}{\mbox{\scriptsize data}}} \:-\:1 \; = \;
\dst\frac{(2.232\pm0.011)\times10^{-3}}{(2.221\pm0.011)\times10^{-3}} \:-\:1
\; \simeq \; 4.95\times10^{-3} \; .
\label{etaexperiment}
\ee

Note, too, that the indirect CPV scale \cite{PLB667p1:qllsm} 
\be
|\varepsilon| \; = \; \frac{2\eta_{+-}+\eta_{00}}{3} \; = \;
(2.228\pm0.011)\times10^{-3}
\label{indirectcpv}
\ee
is quite near the $\alpha/\pi$ scale in Eq.~(\ref{alphapi}), and about
one-half the SQL scales in Eqs.~(\ref{etatheory},\ref{etaexperiment}).

Other measures of CPV are the phase angles $\delta_L$ (also called $A_L$)
and $\phi_{+-}$, related by
\cite{PRD53p2421:qllsm,CMPPNPC2p241:qllsm}
\be
\delta_L \; = \; 2|\varepsilon|\cos\phi_{+-} \; \simeq \; 3.232\times10^{-3}
\label{deltaepsilonphi}
\ee
via Eq.~(\ref{indirectcpv}), and by the observed \cite{PLB667p1:qllsm} angle
$\phi_{+-}=(43.51\pm0.05)^\circ$. The latter value of $\phi_{+-}$ is near
\cite{MRR69p641:qllsm}
\be
\phi_{+-} \; = \; \arctan\left[\frac{2\Delta m_{LS}}{\Gamma_{K_S}}\right]
\; \simeq \; 43.47^\circ \; ,
\label{phils}
\ee
extracted from \cite{PLB667p1:qllsm} $\Delta m_{LS}=3.484\times10^{-12}$~MeV
 and 
$\Gamma_{K_S}=\hbar/\tau_{K_S}=7.351\times10^{-12}$~MeV. Lastly, another
measure of the CPV phase angle $\delta_L$ is the rate asymmetry in semileptonic
weak decays \cite{PLB667p1:qllsm},
i.e.,
\be
\delta_L \; = \; \frac
{\Gamma_{K_L\to\pi^-\ell^+\nu}\,-\,\Gamma_{K_L\to\pi^+\ell^-\bar{\nu}}}
{\Gamma_{K_L\to\pi^-\ell^+\nu}\,+\,\Gamma_{K_L\to\pi^+\ell^-\bar{\nu}}}
\; = \; (3.32\pm0.06)\times10^{-3} \; ,
\label{delta}
\ee
which is not far from the value in Eq.~(\ref{deltaepsilonphi}).

Now we look at a possible source of CPV in the context of the 
SM Cabibbo-Kobayashi-Maskawa (CKM)
\cite{PRL10p531:qllsm,PRL10p531_2:qllsm,PRL10p531_3:qllsm,PTP49p652:qllsm}
matrix, which for convenience we write in the original parametrization due to
Kobayashi and Maskawa \cite{PTP49p652:qllsm} (see also the PDG
\cite{PLB667p1:qllsm} CKM review), viz.\
\cite{PRD53p2421:qllsm,MPLA19p2267:qllsm}
\be
V \; = \; \left(\!\begin{array}{ccc}
V_{ud} & V_{us} & V_{ub}\\ V_{cd} & V_{cs} & V_{cb}\\ V_{td} & V_{ts} & V_{tb}
\end{array}\!\right) \; \to \; 
\left(\!\begin{array}{ccc}
c_1 & -s_1 & \!\!\!0 \\ s_1 & c_1 & \!\!\!0 \\ 0 & 0 & \!\!\!-c_2c_3(1+i\delta)
\end{array} \!\right) \; .
\label{ckmc}
\ee
Here, we have introduced a small CPV complex phase $\delta$ by writing
the almost unity $V_{tb}$ element as
$-c_2c_3e^{i\delta}\simeq-c_2c_3(1+i\delta)$, in the limit of
a real $SU(4)$ Cabibbo submatrix, with $\theta_1=-\theta_C$ and
$\theta_2,\theta_3\to0$ \cite{PRD53p2421:qllsm,MPLA19p2267:qllsm}. An SM
 mechanism that
may give rise to such a complex phase is a nonstandard $WW\gamma$ vertex of
the form \cite{PRD33p3449:qllsm,PRD53p2421:qllsm,MPLA19p2267:qllsm}
\be
\left<\gamma_\mu(q)W_\beta|W_\alpha\right> \; = \;
ie\lambda_w\,\epsilon_{\alpha\beta\mu\sigma}\,q^\sigma \; ,
\label{wwgamma}
\ee
contributing to the tree-level and loop-order $t\to b\,W^+$ mixing graphs of
Fig.~\ref{tbw}.
\begin{vchfigure}[h]
\begin{tabular}{ccc}
\includegraphics[scale=0.6]{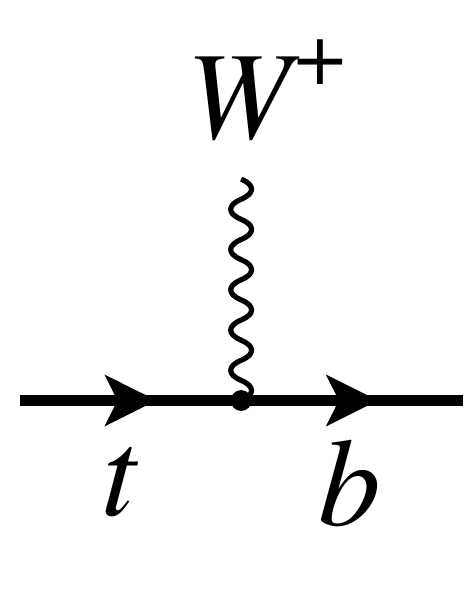} &
\hspace*{6mm}
\includegraphics[scale=0.6]{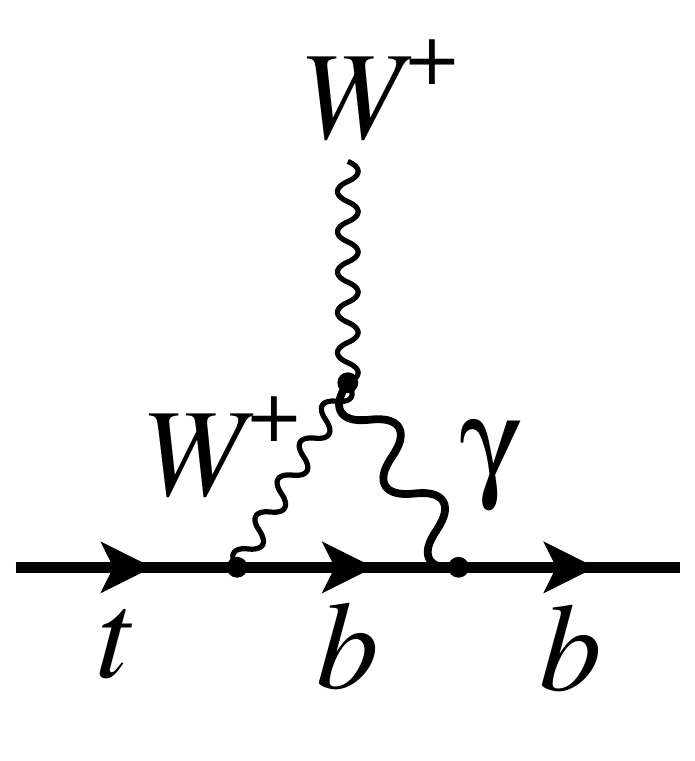} 
\hspace*{6mm}
\includegraphics[scale=0.6]{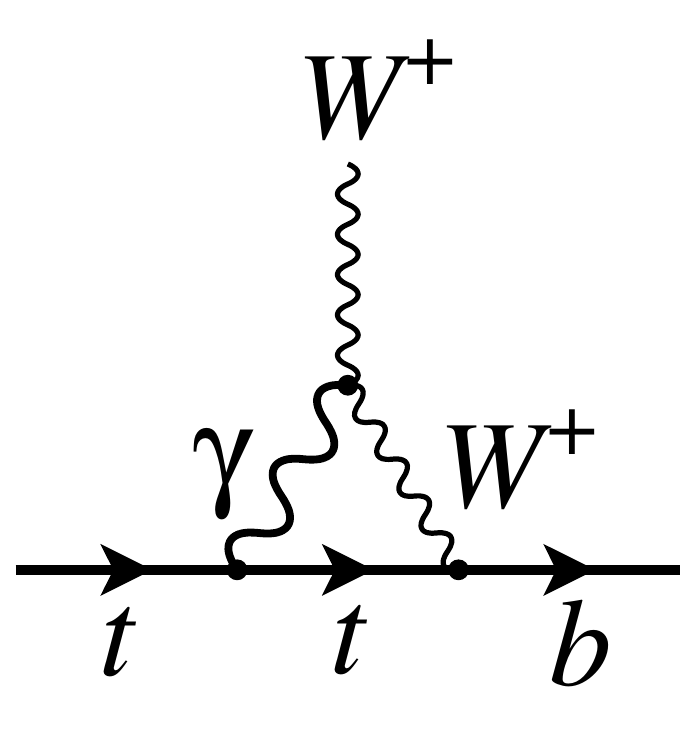} &
\end{tabular}
\mbox{ } \\[-6mm]
\vchcaption{$t\to bW^+$ transition. Left: tree graph; middle and right:
nonstandard $W\gamma W$ loop graphs.}
\label{tbw}
\end{vchfigure}
Evaluation of these graphs yields \cite{PRD53p2421:qllsm,MPLA19p2267:qllsm}
\be
-V_{tb} \; \simeq \; 1+i\lambda_w\frac{\alpha}{\pi}
\ln\left(1+\frac{\Lambda^2}{m_t^2}\right) \; ,
\label{vtb}
\ee
where $\Lambda$ is an ultraviolet (chiral) cutoff. If we assume that 
this cutoff is given by the Higgs mass, Eqs.~(\ref{ckmc},\ref{vtb}) lead to
\be
\ln\left(1+\frac{m_H^2}{m_t^2}\right) \; = \; c_2c_3\,\frac{\pi}{\alpha}\,
\delta \; \simeq \; 1.428 \; ,
\label{higgscutoff}
\ee
where we have used $c_2c_3=0.99913$ \cite{PLB667p1:qllsm}, and
$\delta\simeq3.32\times10^{-3}$ from Eq.~(\ref{delta}).
Substituting now the predicted $m_t\simeq176.8$~GeV from Eq.~(\ref{mt}),
we obtain
\be
m_H\;\simeq\;\sqrt{e^{1.428}-1}\,m_t \; \simeq \; 314.8\;\mbox{GeV} \; ,
\label{mhiggscutoff}
\ee
which is very close to the EWSM value of 314.9~GeV in Eq.~(\ref{mhiggsewsm}),
and also reasonably near the KSRF value of 322~GeV,
both predicted in Sec.~\ref{tophiggs} above (also see
\cite{JPG32p735:qllsm}). Moreover, all three NP Higgs scales are
roughly  compatible with the observed \cite{PLB667p1:qllsm}
$m_t=(173.5\pm0.6\pm0.8)$~GeV as well.

In conclusion, we should emphasize that our \em nonperturbative Standard-Model
\em \/scenarios for CPV and the Higgs are in no way related to the usual \em
perturbative field-theory approach beyond the SM. \em Another NP approach to
CPV in the SM can be found in \cite{HEPPH0308030:qllsm}.

\section{Concluding remarks}

In this review article we have shown how a huge number of strong,
electromagnetic, and weak processes fit very well within a scheme whereby
chiral symmetry is spontaneously generated in a {\em linear} \/representation.
The pion, sigma meson, and their $SU(3)$ partners arise dynamically through the
quark interactions, and they govern a large body of data with very few
parameters indeed. The only {\em small} \/corrections to the chiral limit are
due to the current quark masses of the lightest quarks ($u$, $d$, $s$).

Sections 1--15 above covered the many facets of this description. One
might think of extrapolating the ideas to the top-quark sector as we did in
Sects.~16 and 17, but this is more speculative and perhaps also more
problematic because one is moving a fair way from the chiral (zero-mass) limit.
An attempt along such lines is fraught with difficulties since the
chiral-symmetry-breaking corrections are surely more substantial.

Our description of the Higgs boson as primarily a $t \bar{t}$ composite yielded
a mass of about 315 GeV. As mentioned in Sec.~16, this is far removed from the
preferred values resulting from fits to electroweak data, though these limits
should be interpreted with some caution 
\cite{MPLA10p845:qllsm,MPLA10p845_2:qllsm,MPLA10p845_3:qllsm}.
We are also aware of the recent Higgs-like signals at the LHC,
observed by the ATLAS collaboration \cite{PLB716p1:qllsm} at 126~GeV
and by the CMS collaboration \cite{PLB716p30:qllsm} at 125~GeV. However,
insufficient statistics has not allowed so far to pin down the spin-parity of
the found state, which could be a scalar, pseudoscalar, or a tensor boson, in
view of the seen two-photon decay mode. Thus, one cannot exclude yet an
interpretation of the data by e.g.\ a technipion, as predicted in certain
(extended) technicolor models, which will have very similar decay modes
\cite{PRD33p93:qllsm,PRD33p93_2:qllsm,PRD33p93_3:qllsm}, albeit with different
angular distributions. Alternatively, the enhancement at 125 GeV might be just
one of several threshold enhancements due to a possible substructure in the
weak-interaction sector \cite{ARXIV13047711:qllsm}. So we await
with considerable interest the high-statistics measurements to be done after
the LHC ugrade in a couple of years, which will hopefully allow to carry out
the required partial-wave analyses of the observed boson's decay products.
Nevertheless, irrespective of the possible confirmation of a Higgs-like scalar
at about 125~GeV, the existence of another scalar with a mass of roughly
320~GeV is not out of the question.

Despite the power and simplicity of the \qllsm, other, more traditional
approaches to the \lsm\ have been appearing in the literature, even very
recently. For instance, in \cite{PRD82p054024:qllsm} a \lsm\ with only
mesonic degrees of freedom and global chiral symmetry was formulated, whose
Lagrangian contains elementary vector and axial-vector fields, besides the
usual pseudoscalar and scalar ones. Note that this is a very complicated,
perturbative model with several adjustable parameters, but in principle
applicable to
both the light scalar nonet and the scalars in the energy region
1.3--1.7~GeV. More relevant for the present NP theory is a formal
selfconsistent generalization \cite{AIP1030p412:qllsm,AIP1030p412_2:qllsm}
of the \qllsm\  by
dynamical generation beyond one loop, including sunset-type diagrams.
Such an approach, based on the imposed exact cancellation of quadratic
divergences, may eventually allow an asymptotically free formulation of
strong interactions, as an effective alternative \cite{AIP1030p412:qllsm} to
 QCD, also at higher energies.
An asymptotically free \qllsm\ belongs \cite{CJP55p1123:qllsm} to a class of
non-Hermitian yet $PT$-symmetric field theories, which have been actively
pursued by especially C.~Bender \cite{PRL80p5243:qllsm} and collaborators.

As a final remark, let us stress once again the importance of the NP nature
of the \qllsm, as formulated in the present review. Due to the applied
bootstrap principle, all couplings of the theory are selfconsistently
interrelated via dynamical generation and loop shrinking
 \cite{MPLA10p251:qllsm}, leaving no model parameters, save the ---
 experimentally fixed \cite{PLB667p1:qllsm} --- pion weak decay constant
 $f_\pi$. In spite of this lack
of freedom, or more likely thanks to it, a wealth of low-energy observables
can be described, some even with amazing precision. No other effective theory
of strong interactions comes even close in performance.

\begin{acknowledgement}
One of us (MDS) is deeply indebted to his former co-authors A.~Bramon,
V.~Elias,
N.~H.~Fuchs,
H.~F.~Jones, and N.~Paver for their valuable
contributions to the development of the \qllsm, and also to E.~van Beveren,
F.~Kleefeld, and many others (see references below) for collaboration on
several related applications.
The figures in this review have been produced with the graphical package 
{\sl SCRIBBLE} \cite{NIMA389p305:qllsm}, and we thank P.~Nogueira for helpful
suggestions.
This work was supported by the {\it Funda\c{c}\~{a}o para a Ci\^{e}ncia e a
Tecnologia} \/of the {\it Mi\-nist\'{e}rio da Ci\^{e}ncia, Tecnologia e Ensino
Superior} \/of Portugal, under contracts nos.\ CERN/FP/\-83502/\-2008 and
CERN/FP/109307/2009.
\end{acknowledgement}

%
%

\bibliographystyle{fdp}
\bibliography{qllsm}
\end{document}